\documentclass{article}

\usepackage[english]{babel}

\usepackage[letterpaper,top=2cm,bottom=2cm,left=3cm,right=3cm,marginparwidth=1.75cm]{geometry}

\usepackage{amssymb}
\usepackage{float}
\usepackage{booktabs}
\usepackage{amsmath}
\usepackage{graphicx}
\usepackage{subcaption}
\usepackage{afterpage}
\usepackage[flushleft]{threeparttable}
\usepackage[hyphens,spaces,obeyspaces]{url}
\usepackage[colorlinks=true, allcolors=blue, breaklinks=true]{hyperref}
\usepackage{multirow}
\usepackage{array}
\usepackage{tabularx}
\usepackage{threeparttable}
\usepackage[square,numbers]{natbib}

\begin{document}

\begin{titlepage}

    \centering
    \includegraphics[width=0.5\textwidth]{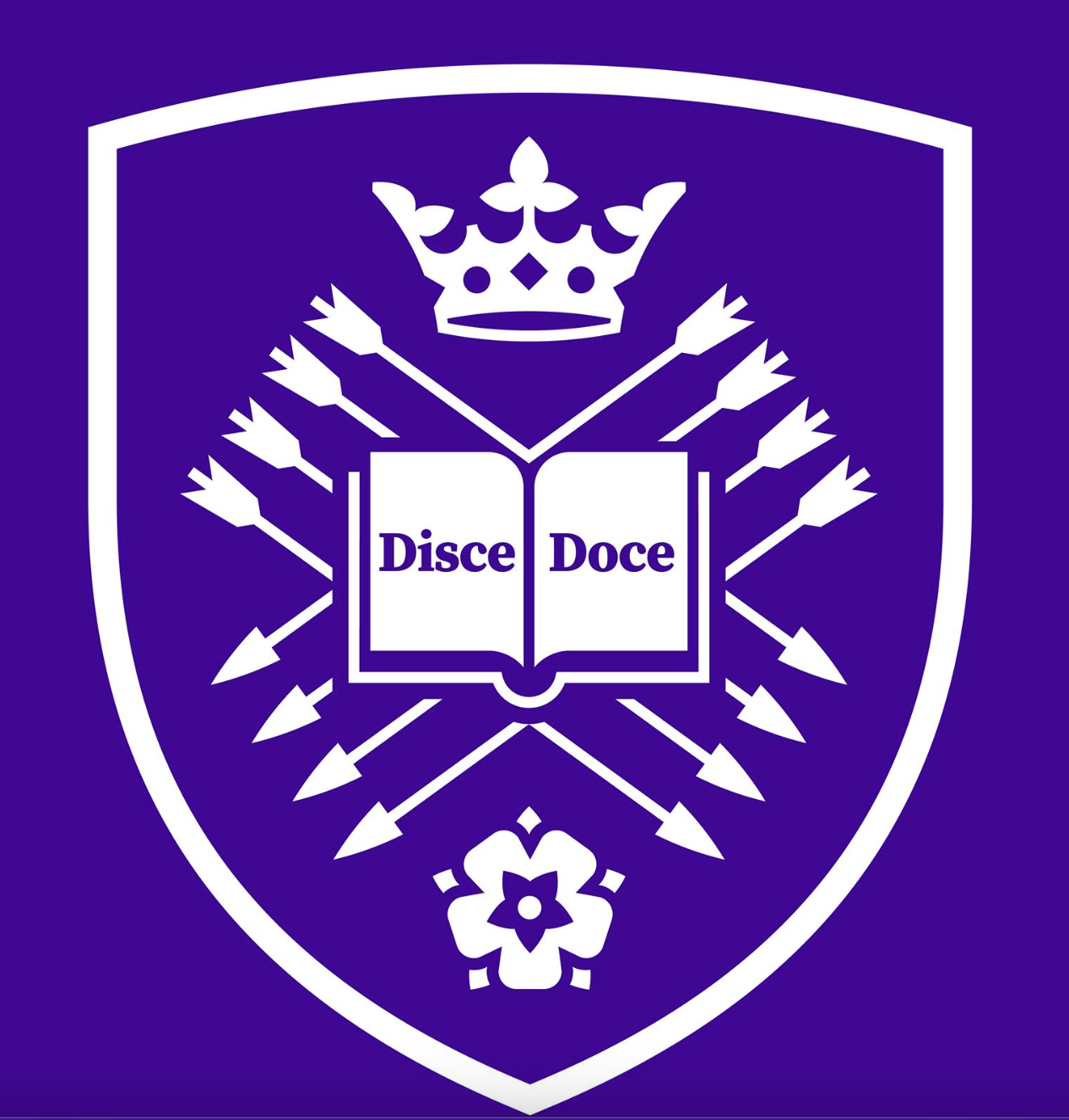}\par\vspace{1cm} % Replace with the path to your logo file
    {\scshape\LARGE University of Sheffield \par}
    \vspace{1.2cm}
    {\huge\bfseries Zero-Dimensional Cardiovascular Modeling:\\ A Personalized Approach to Non-Invasive Measurement and Sensitivity Analysis \par}
    \vspace{0.5cm}
    {\Large Supervisor: Dr. Xu Xu\par} 
    \vspace{0.5cm}
    {\itshape A report submitted in fulfilment of the requirements for the degree of \\ MSc in Data Analytics \\ in the \\ Department of Computer Science\par}
    \vspace{0.5cm}
    {\Large\bfseries By Team Kilo \par}
    \begin{tabular}{c} % Center-align the text within the tabular environment
        Akio Nishida \\
        Bhagyashree \\
        Jiacheng Liu \\
        Pranav Kumar Sasikumar \\
        Simran Wadhwa \\
        Wendi Jiang \\
    \end{tabular}
    \vfill
    \vspace{0.5cm}
    {\large \today\par}

\end{titlepage}

\newpage
\tableofcontents
\newpage
\listoftables
\listoffigures

\newpage
\begin{abstract}
Zero-dimensional cardiovascular models provide a computationally efficient framework for studying global hemodynamic behavior, yet the influence of model complexity on parameter sensitivity remains insufficiently understood. This work investigates two lumped-parameter cardiovascular models, a simplified single-ventricle configuration and a detailed four-chamber representation, to examine how physiological parameter sensitivities vary with model structure. Time-varying elastance functions are used to represent cardiac dynamics, and global sensitivity analysis is performed using Sobol and Morris methods to quantify the impact of key physiological parameters, including venous return, myocardial contractility, total peripheral resistance, and arterial compliance. The results demonstrate that sensitivity rankings differ substantially between the two models, highlighting the role of model granularity and parameter interactions in shaping cardiovascular responses. These findings support sensitivity-driven model reduction and provide a foundation for scalable, non-invasive cardiovascular simulation frameworks.
\end{abstract}

\newpage
\section{Executive Summary}

This project explores the use of zero-dimensional (0D) modelling techniques to analyse cardiovascular responses. Its goal is to understand how different physiological parameters affect cardiovascular function. A comparative analysis was performed on two separate cardiovascular 0D models: one simulating a solitary ventricle and the other representing a four-chamber configuration. Our hypothesis suggested that these models would produce distinct sensitivity outcomes as a result of their different levels of intricacy, which encompass model granularity, parameter inclusion, and interaction effects. It's important to note that the sensitivity rankings are contingent upon the available output measurements.
\\ \\
\textbf{Experimental Design:} The study utilised a combination of 0D modelling and sensitivity analysis techniques. By employing advanced techniques such as Sobol and Morris methods, sensitivity analysis enabled us to methodically assess the influence of different physiological parameters on simulated cardiovascular responses. Our objective was to identify the crucial factors that have a significant impact on cardiovascular dynamics by strategically manipulating these parameters.
\\ \\
\textbf{Model Development:} Two separate 0D models were examined to depict the cardiovascular system, and one model was derived from the second model:

\begin{itemize}
\setlength\itemsep{-0.3em} % Reduces the space; adjust the value as needed
    \item The \textbf{Single Ventricle Model} is a simplified representation of the heart that combines all its components into a single pumping unit. It includes important factors such as venous capacitance, arterial compliance, and peripheral resistance to accurately depict the dynamics of blood flow.
    \item The \textbf{Four-Chamber Model} is a more intricate representation that explicitly depicts the four chambers of the heart, namely the right atrium, right ventricle, left atrium, and left ventricle. Each chamber was designed as an individual compartment with unique pressure-volume characteristics and valve behaviour.
\end{itemize}

\textbf{Sensitivity analysis }was conducted to evaluate the influence of different physiological parameters on the simulated cardiovascular responses. The key parameters that were analysed included venous return, myocardial contractility, total peripheral resistance, and arterial compliance. We deliberately manipulated these parameters within the normal range of physiological values in order to determine the crucial factors that have a significant impact on the simulated responses. From the results obtained from the Sobol Method and the Morris Method applied to the first model, our objective is to obtain valuable insights into:

\begin{itemize}
\setlength\itemsep{-0.3em} % Reduces the space; adjust the value as needed
    \item The impact of the level of model detail on the sensitivity to specific physiological parameters.
    \item Can simpler models, obtained by fixing unimportant parameters in the complex model, achieve comparable results while being more computationally efficient?
\end{itemize}

The selection of parameters to be fixed in the third model will be determined by the sensitivity analysis of the second model. Parameters that have been determined to have minimal or well-established impact on cardiovascular responses can be given fixed values, which can potentially make the model simpler without significantly affecting its accuracy.
\\ \\
\textbf{Results and Conclusions:} Both 0D models showed the expected hemodynamic responses—elevated arterial blood pressure, increased myocardial contractility, and increased venous return—under simulated conditions. However, the sensitivity analysis showed significant differences between the four-chamber and single ventricle models, emphasising the importance of choosing a model and complexity level that matches the research question. Future simulations and sensitivity analysis would benefit from adding physiological parameters.\\
Access the Code: The complete code used for the simulations and analysis in this study is available at \href{https://github.com/pranav170620/Sensitivity-Analysis-of-Cardiovascular-Models}{GitHub Repository}.
\\ \\
\textbf{Novel Contributions to Existing Knowledge:}
This study provides new insights into cardiovascular model sensitivity by comparing a single ventricle model to a four-chamber model, highlighting how model complexity impacts sensitivity analysis outcomes. Addressing the gap in previous research that focused on single models, it applies Sobol and Morris global sensitivity analysis methods to both models. This approach offers a novel perspective on reducing unnecessary clinical measurements to lessen patient burden.

\newpage
\section{Introduction}
The human cardiovascular system is an exceptional example of biological engineering, coordinating the circulation of blood and the transportation of oxygen throughout the body. This complex system depends on a fragile equilibrium among different physiological parameters. Acquiring a thorough comprehension of these factors and how they interact is crucial for devising effective approaches to control cardiovascular well-being. Cardiovascular diseases (CVDs) are a prominent cause of mortality globally, responsible for approximately 18.6 million deaths each year, as reported by the British Heart Foundation. Approximately 7.6 million individuals in the United Kingdom are currently afflicted with heart and circulatory ailments. These statistics highlight the importance of improving our knowledge of cardiovascular function in order to effectively address the widespread occurrence and consequences of cardiovascular diseases.
\\
\\
This project investigates the fascinating field of zero-dimensional (0D) modelling techniques for simulating cardiovascular circulation. Computational models are powerful tools for investigating complex physiological mechanisms and predicting how the heart will respond in various situations.\\
\\
By employing sensitivity analysis, specifically, the objective is to elucidate the intricate interplay among physiological parameters within the cardiovascular system.
Sensitivity analysis is an effective method for evaluating the impact of variations in the input parameters of a model on its output. To determine the parameters that have the greatest influence on cardiovascular function, we can systematically manipulate them and observe the resulting changes in the model's predictions. \\
\\
Previous studies have mainly concentrated on cardiovascular modelling, but our knowledge of how these models are affected by specific physiological parameters is still limited. This project seeks to close this gap by employing a comparative methodology. We aim to gain valuable insights into the important physiological parameters that affect cardiovascular behaviour by using different 0D models. Additionally, we will analyse and compare the results. This sensitivity analysis is the first step towards achieving personalised cardiovascular healthcare.\\
\\
This study has the capacity to enhance our comprehension of cardiovascular circulation while simultaneously laying the groundwork for the creation of individualised strategies for managing cardiovascular health. These strategies may be advantageous for individuals with particular health conditions or undergoing invasive medical procedures that impact the cardiovascular system. \\
\\
In addition, the study will present a third model that seeks to streamline the approach while preserving the same level of detail. This streamlined model is designed to decrease computational requirements, making it suitable for simulating real-world situations encountered in clinical settings. \\
\\
This introduction establishes the context for the following sections, which will provide detailed explanations of the specific goals, experimental approaches, and significant discoveries of the project.

\begin{itemize}
    \item \textbf { Background and Literature Review}: This section delves into the general principles of cardiovascular models, detailing zero-dimensional (0D), one-dimensional (1D), and two-dimensional (2D) and three-dimensional (3D) models. It includes a specific review of 0D models in cardiovascular research, highlighting their suitability and limitations. Additionally, it covers sensitivity analysis, clinical measurements, and the research question addressed by this project.
\end{itemize}
\begin{itemize}
  \item \textbf{Methodology}: This section provides a detailed explanation of the specific methods utilized in the investigation. It includes:
    \begin{itemize}
      \item Replication of current 0D models, including both single ventricle and four-chamber models.
      \item Implementation of sensitivity analysis on each model to assess their performance.
      \item Development and explanation of a third, simplified model, along with the reasoning behind its creation.
      \item Justification for selected input parameter values.
    \end{itemize}
\end{itemize}

\begin{itemize}
  \item \textbf{Results}: This section provides an overview of the study’s results, presenting and discussing the sensitivity analyses conducted on each model. It includes:
    \begin{itemize}
      \item Model 1 Sensitivity Analysis and Validation 1.
      \item Model 2 Sensitivity Analysis and Validation 2.
      \item Model 3 Sensitivity Analysis.
    \end{itemize}
\end{itemize}

\begin{itemize}
  \item \textbf{Discussion and Conclusions}: This section integrates all the information and draws final inferences. It analyzes the results and identifies the key physiological parameters that significantly impact cardiovascular behavior using sensitivity analysis. It also examines the consequences of the findings for cardiovascular regulation and personalized health management strategies. It includes:
    \begin{itemize}
      \item Models and Results.
      \item Validation according to Non-Invasive Measurements (5.1.1 and 5.2.1).
      \item Future Improvement.
    \end{itemize}
\end{itemize}

\subsubsection*{Future Research: Prospective Avenues}

As this research advances, numerous intriguing possibilities arise for additional investigation:
\begin{itemize}
    \item  \textit{\textbf{Including supplementary physiological parameters:}} Incorporating variables such as heart rate, blood viscosity, and neurohumoral regulation into the model can enhance the accuracy of the simulations and offer a more comprehensive understanding of cardiovascular function.
    \item  \textbf{\textit{Creation of other Streamlined Model:}} Utilising the discoveries from the four-chamber model, several novel model can be constructed with predetermined values for specific parameters in order to decrease computational requirements. This streamlined model could subsequently be employed in clinical settings for pragmatic applications.
    \item  \textbf{\textit{Experimental Data Validation:}} The inclusion of experimental data in the model validation process can improve the accuracy and dependability of the model. Validating the model's predictive capabilities and ensuring its applicability in clinical scenarios can be achieved by comparing simulated results with real-world data.
\end{itemize}

\section{Background and Literature Review}

\subsection{General}

Cardiovascular models have progressed from basic analogs to sophisticated computational simulations that can accurately replicate complex physiological behaviors. These models are important in predicting cardiovascular responses under different physiological scenarios, where traditional experimental approaches may be limited \cite{eck2016guide}. Zero-dimensional (0D) modeling is a computationally efficient framework that reduces the cardiovascular system to a set of lumped parameters. These parameters describe system dynamics over time using ordinary differential equations, allowing for quick iterations and adaptations to different physiological conditions without the computational burden of more spatially detailed models.

\subsection{Model}

A variety of approaches to computational modeling can be used depending on the study's objectives and constraints. These approaches range from zero-dimensional (0D) models to more complex three-dimensional (3D) simulations, each with distinct advantages and applications.

\subsubsection{Zero-dimensional (0D) models}
Zero-dimensional models, by definition, lack spatial dimensions and instead focus solely on system dynamics over time. This simplification is especially useful when detailed spatial resolution is not required for the analysis's objectives or when computational resources are limited. The primary advantage of using 0D models is their low computational demand. These models reduce complex physical systems to time-dependent ordinary differential equations (ODEs) without spatial variation, simplifying the PDE system \cite{esmaily2013}. This reduction in computational requirements enables faster simulations and the ability to perform multiple iterations under different scenarios, making 0D models ideal for exploratory studies and sensitivity analysis. Furthermore, 0D models enable a better understanding of systemic relationships and fundamental dynamics, which is essential for identifying key drivers and interactions within the system \cite{pfaller2022}.

\subsubsection{One-dimensional (1D) models}
One-dimensional models strike a balance between computational efficiency and spatial resolution, depicting the cardiovascular system as a network of interconnected tubes with flow and pressure changes considered along a single spatial dimension. These models are especially useful for simulating blood flow in large vessels, where changes occur along the length but can be averaged over the cross-sectional area \cite{formaggia2010cardiovascular}. While more computationally intensive than 0D models, 1D models provide detailed insights into wave propagation and pressure dynamics, which are critical for studies on arterial stiffness and pulse wave velocity.

\subsubsection{Two-Dimensional (2D) and Three-Dimensional (3D) Models} These models provide the most detailed representation of cardiovascular structures. These models are extensively used for detailed studies of complex flow patterns and mechanical stresses within the heart and major vessels \cite{taylor1998finite}. They require significant computational resources and are typically used when precise spatial resolution is required, such as in surgical intervention planning or detailed blood flow analysis in congenital heart diseases. However, the high computational cost and complexity of creating accurate 3D geometries can make them unsuitable for routine or large-scale studies.

\subsubsection{Review of Zero-Dimensional (0D) Models in Cardiovascular Research} These models view the cardiovascular system as a network of lumped parameter elements, with each element representing a different aspect of the system, such as compliance, resistance, or inertance. Notable examples and a critical assessment of their methods are:

\begin{itemize}
    \item \textbf{Windkessel Models}: Represent the arterial system using resistive and capacitive elements. Extensively used, but cannot capture waves.  
 \cite{westerhof2009arterial}.
    \item \textbf{CircAdapt Model}: Integrates heart and vasculature into one framework, effective for studying disease progression, but limited by adaptability assumptions.
 \cite{arts2005adaptation}.
    \item \textbf{Lumped Parameter Heart Models}: Incorporate detailed valve dynamics, useful for simulating clinical conditions, but difficult to calibrate due to complex parameter interdependencies. \cite{mynard2015one}.
\end{itemize} 
\

\subsubsection{Suitability and Limitations}
The suitability of 0D models is primarily determined by their ability to provide rapid simulations and adapt to various physiological and pathological conditions while requiring minimal computational resources. They are well-suited for educational purposes, preliminary medical device design, and sensitivity analyses. However, their limitations include a lack of spatial resolution, oversimplification of physiological interactions, and a reliance on empirical data for parameter estimation.

\subsubsection{Conclusion}
We will use zero-dimensional models for cardiovascular circulation simulation in the upcoming study because they are computationally efficient and adaptable. These models are appropriate for our research objectives, as they concentrate on systemic interactions and parameter sensitivity rather than localized physical phenomena.

\subsection{Sensitivity Analysis}

Sensitivity analysis is broadly categorised into two types: local and global. Local sensitivity analysis focuses on the effects of slight variations in parameters around specific values, detailing the immediate model response at such baseline. Conversely, global sensitivity analysis, which we employ exclusively in our project, investigates how changes across the entire range of input parameters impact a model's output. This comprehensive method is essential for understanding complex cardiovascular models, as it thoroughly explores parameter spaces and provides robust insights into both direct impacts and interactions among parameters. Particularly useful in models with significant uncertainties and nonlinear behaviours, global sensitivity analysis enhances decision-making by identifying key influencers and optimising resource distribution in research and development. By leveraging variance-based techniques like Sobol indices, our approach not only improves the reliability of the model but also supports informed strategic decisions crucial for managing different heart conditions.
\\ \\
Weighted average of local sensitivity analysis (WALS), partial rank correlation coefficient (PRCC) and Multi-parametric sensitivity analysis (MPSA) are efficient in computation but have limitations of use depending on the model structure and cannot extract higher order of interaction of parameters. \cite{zhang2015} Unlike the commonly used methods, the Sobol method can calculate the effect precisely even if the model is nonlinear and displays both the first order and total impacts. These strengths come from its formula, which shows that the output variance can be decomposed into each parameter’s contribution. In fact, this method is used to perform sensitivity analyses of models on parts of the circulatory system. \cite{pathmanathan2019}, \cite{gul2015} It can be considered that these studies commonly compute the impact of individual parameters with high preciseness regardless of the number of parameters and outputs and the model complexity and clearly show the influential parameters. Therefore, the Sobol method is rational for this research because of its reliability and ease of interpreting the results. Expanded Fourier amplitude sensitivity analysis (eFAST) is another alternative method commonly used in clinical sensitivity analysis. \cite{rabiee2023} Based on the results of using eFAST for medical-related analysis, this method is as reliable as the Sobol method and more efficient in computation than that. However, eFAST was not adopted in our study, considering the time limitation of this project and its theoretical complexity compared to the Sobol method.
\\ \\
However, the disadvantage of the Sobol method is its high computational costs \cite{zhang2015}, which may not be suitable for specific patients requiring emergency treatment, such as patients with sepsis. On the other hand, the Morris method is cost-efficient and well-suited when the model contains little uncertainty. \cite{xu2019} The Morris method is used in a study to reproduce left ventricular pressure and volume from patient-specific data. \cite{taconné2021} While the Morris method is less reliable in identifying the parameter impact than the Sobol method, the Morris may be sufficient for even other complex cardiovascular systems to adopt and worthwhile examining further based on the fact that the parameter values estimated from the results of this sensitivity analysis were almost identical to the experimental values. Hence, we also employ this method in our first model to briefly compare its result with the result using the Sobol method.

\subsection{Clinical Measurements}

After estimating each parameter's impact through sensitivity analysis in a specific model, the actual values of the inputs can be identified by solving the inverse problem. This approach is applied to the personalisation process in clinical fields. \cite{saxton2024} In other words, the sensitivity analysis results using a model that accurately represents physiological motions enable the identification of individual patients' specific characteristics and provide optimal diagnostic and treatment strategies. \cite{volzke2013}
\\ \\
However, measuring all outputs from sensitivity analysis is not always feasible from the perspective of various components, such as the risks to patients and technical limitations. For example, invasive cardiac catheterisation can directly measure cardiovascular conditions such as the blood pressure in each heart chamber by inserting a catheter into an artery or vein in a particular body part. Nevertheless, this invasive approach bruises the patient's body part in which the catheter was inserted or in rare cases, it may also cause heart attack and stroke as the blood supply is interrupted. \cite{cascino2023} , \cite{nhs2022} It is ideal to acquire valid sensitivity analysis results even when relying solely on non-invasive measured outputs to minimise patients' risk while considering the equipment costs for non-invasive measurements and the complexity of measurement procedures. Therefore, comparing the results of a sensitivity analysis using all outputs with the results of a sensitivity analysis using only non-invasive measurement values allows further discussion in this study on to what extent the model we use is applicable in an actual clinical environment.
\\ \\
The left ventricular pressure is important for the study of cardiac mechanics and those values is typically obtained through catheterization. \cite{suga1990} Although there are some recent researches to identify those values through non-invasive method  \cite{arvidsson2023}, it can be considered that the method of measuring them has not been established. The pressure in the systemic bloods can be regarded as the other outputs that cannot be measured by reliable non-invasive measurements. While the cuff/stethoscope method is typically used for measuring arterial blood pressure, it differs considerably from the intra-arterial pressure.  \cite{meidert2018} Therefore, it is inavoidable to measure the value invasively even if the cost of equipping cuffs and the damage to the body caused by invasive measurements are considered. On the other hand, the volume of each compartment of the heart is established to measured by echocardiography.  \cite{lang2015} Based on the results of the previous studies, it is valid to determine these values using echocardiography from the perspective of its accuracy. 

\subsection{Research question}

In this study, in simulations of the cardiovascular system using zero-dimensional compartmental cardiovascular models, sensitivity analyses were used to assess and determine the importance of model input parameters and to guide the individualised setting of model parameters. Further, these models compare how the effects of input parameters on model predictions differ in different cardiac structural complexities (e.g.single ventricle models versus four-chamber heart models). Additionally, using the kind and availability of clinical output data to optimize sensitivity indices and parameter rankings improves the models' predictive accuracy and clinical applicability.
\\ \\
Throughout the investigation, the dynamic response of the cardiovascular system is simulated using a zero-dimensional (0D) modeling technique, which reduced the intricate structure of the cardiovascular system to circuit components with fluid dynamics, such as resistors, capacitors, and inductors. This study compared two alternative cardiovascular models, one with a single ventricle and the other with a four-chamber heart, to see how the model architectures differed in terms of how accurate the simulation findings were.
\\ \\
Furthermore, This study will use sensitivity analysis to determine which cardiovascular model parameters most affect the simulation results' accuracy. This is an important step towards optimizing model parameters and directing customized settings. This investigation will further our comprehension of the mechanisms controlling the cardiovascular system in various physiological contexts. The study will also explore how clinical data can be used to optimise sensitivity indices and parameter rankings, thereby improving the predictive accuracy and clinical applicability of models. These efforts will promote the scientific understanding and practical application of models to support more effective treatment strategies for cardiovascular diseases.

\section{Methodology}

\subsection{Model Replication}

The methodology will be structured into three main parts. Initially, the goal would be replicating two models of varying complexity, making necessary adjustments to align them with our objectives. Next, we will conduct sensitivity analysis using the Sobol Method on both models under specific conditions to assess parameter significance through the Average Direct-Effect Index ($S_1$) and Total-Effect Sobol Sensitivity Indices ($S_T$). Additionally, we will apply the Morris Method to the first model to compare its results and computational efficiency with the Sobol Method. Finally, based on the outcomes from the second model, we will develop a third model. This model will have the same structure as the second but will incorporate fixed values for parameters labelled as unimportant or well-established, thus simplifying the model. \textbf{Most equations and theoretical models discussed in these sections are derived from \cite{saxton2023} for the first model and \cite{colunga2023} for the second model}. Additionally, we would incorporate an additional section explaining the rationale behind selecting suitable input parameter values, as these choices can significantly impact the model outputs.
\\ \\
We will begin by exploring the first model, an advanced simulation of the cardiovascular system that focuses on systemic circulation, with a particular emphasis on the left ventricle. We plan to adhere to the established settings of the Ordinary Differential Equations and parameters, but will also make necessary adjustments based on our design. Here is a short overview of the first model which consists of three compartments representing different parts of the cardiovascular system.
\\ \\
\textbf{Model 1}
\\ 
As the most initial setting of the model, the state of each compartment it concerned of is defined by:
\begin{equation}
    (V_i(t), P_i(t), Q_i(t)) \quad \text{for } i \in \{lv, sa, sv\}
\end{equation}
where $V$ is volume(mL), $P$ is pressure(mmHg), and $Q$ is inlet flow (mL/s) as time-dependent dynamics; and $lv$ stands for left ventricle, $sa$ is systemic arteries, and $sv$ is systemic veins; together as the prior, current, and proceeding compartments of the model. \cite{saxton2023} \\
\\
The equations governing volume and pressure changes are given by:
\begin{align}
    \frac{dV_{s,i}}{dt} &= Q_i - Q_{i+1}, \\
    \frac{dP_i}{dt} &= \frac{1}{C_i} (Q_i - Q_{i+1}), \\
    Q_i &= \frac{P_i - P_{i+1}}{R_i}.
\end{align}
Here $V_{s,i}$ denotes the stressed volume, $C_i$ is the compartmental compliance, and $R_i$ the resistance between compartments i and i+1. \cite{saxton2023}
\\ \\
Also, since the model only concerns about left ventricle and the corresponding arteries and veins, the valve dynamics are modeled using diode-like behavior, allowing flow in one direction with resistance \( R_{val} \) when \( P_i > P_{i+1} \), and blocking it otherwise. Therefore for compartments from $lv$ and $sa$ can be further represented as:
\begin{equation}
    Q_i = 
    \begin{cases} 
        \frac{P_i - P_{i+1}}{R_{val}} & \text{if } P_i > P_{i+1} \\
        0 & \text{if } P_i \leq P_{i+1}
    \end{cases}
\end{equation}
\\ 
When coming to the left ventricle, the model should also consider about the elastance of the Left Ventricle: - The left ventricle's time-varying compliance \( C(t) \), or elastance \( E(t) \), is central to the model's ability to simulate heart dynamics. Elastance, the reciprocal of compliance, is modeled as:
\[ E_{lv}(t) = \frac{P_{lv}(t)}{V_{Stressed}(t) - V_{unstressed}} \]
- \( E(t) \) varies over the cardiac cycle and is described analytically by the expression:
\[ E(t) = (E_{max} - E_{min}) \cdot e(t) + E_{min} \]
where $E_{max}$ and $E_{min}$ are maximal and minimal elastance, and \( e(t) \) is an activation function defined over the cardiac cycle with parameters for end systolic and end pulse times $\tau_{es}$ and $\tau_{ep}$:
\[ e(t) =
\begin{cases}
\frac{1}{2} \left(1 - \cos\left(\frac{\pi t}{\tau_{es}}\right)\right) & \text{if } 0 \leq t < \tau_{es} \\
\frac{1}{2} \left(1 + \cos\left(\frac{\pi (t - \tau_{es})}{\tau_{ep} - \tau_{es}}\right)\right) & \text{if } \tau_{es} \leq t < \tau_{ep} \\
0 & \text{if } \tau_{ep} \leq t < \tau
\end{cases} \]

\begin{figure}[ht]
    \centering
    \begin{subfigure}{0.4\linewidth}
        \centering
        \includegraphics[width=\linewidth]{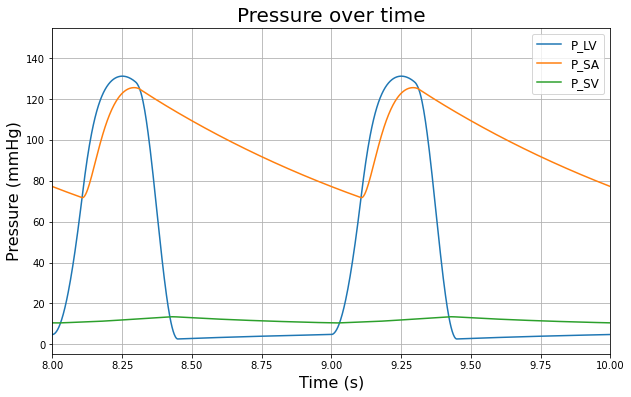}
        \caption{Pressures}
        \label{fig:model1-pressure}
    \end{subfigure}%
    \hspace{0.05\linewidth} 
    \begin{subfigure}{0.4\linewidth}
        \centering
        \includegraphics[width=\linewidth]{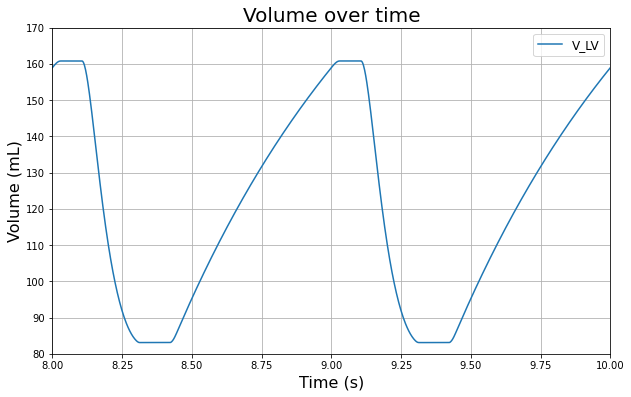}
        \caption{Volume}
        \label{fig:model1-volume}
    \end{subfigure}%
    \hspace{0.05\linewidth}
    \begin{subfigure}{0.4\linewidth}
        \centering
        \includegraphics[width=\linewidth]{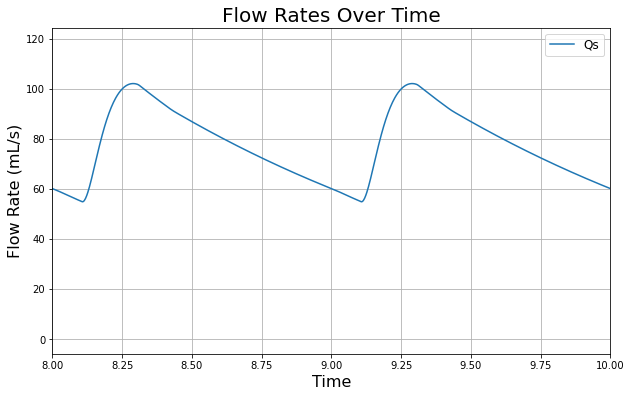}
        \caption{Flow Rate}
        \label{fig:model1-flow}
    \end{subfigure}
    \hspace{0.05\linewidth}
    \begin{subfigure}{0.4\linewidth}
        \centering
        \includegraphics[width=\linewidth]{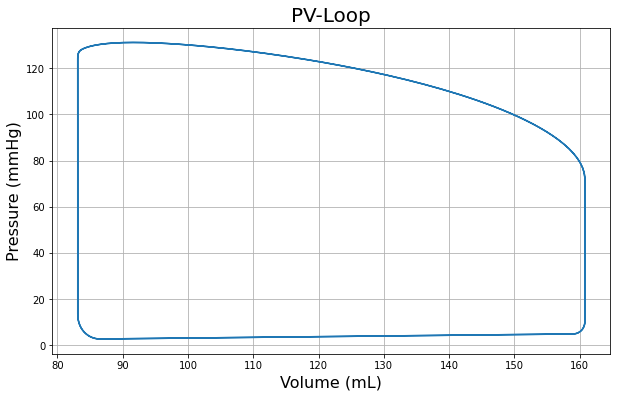}
        \caption{Pressure-Volume Loop of Left Ventricle}
        \label{fig:model1-pvloop}
    \end{subfigure}
    \caption{Model 1 outputs with sub-interval from 8s to 10s. Figure a) illustrates the pressure outputs of the left ventricle as well as the pressures of system arteries and veins. Figure b) and c) are the volume of the left ventricle over time and the flow rate over time respectively. Figure d) contains the Pressure-Volume Loop which compares pressure and volume of left ventricle for each cardiac cycle}
    \label{fig:model1-1to4}
\end{figure}

\cite{saxton2023} and in this case the simplification form according to all of the above equations for change in pressure of the left ventricle would be:

\begin{equation}
    \frac{dP_{lv}(t)}{dt} = E(t)(Q_{lv}(t) - Q_{sa}(t)) + \frac{P_{lv}(t)}{E(t)} \frac{dE(t)}{dt}
\end{equation}
\\
Now we have the necessary equations to analyze all compartments. Our focus will be on examining the three aspects of pressure $P_{i}$, alongside the volume $V_{i}$ and inlet flow $Q_{i}$ in the left ventricle, which serve as the output estimates from the model. The input parameters are all the remaining variables listed directly from the original source in the Table \ref{tab:model1_parameters}. The resulting data is visualised in Figure \ref{fig:model1-1to4}, which seems to successfully replicate the desired result from the original model.
\\

\begin{table}[H]
\centering
\begin{threeparttable}
\caption{Model input parameters and their initial values.}
\label{tab:model1_parameters}
\begin{tabular}{@{}lllr@{}}
\toprule
Parameter ($\theta$) & Unit & Description\\ \midrule
$E_{max}$ & mmHg/ml & Maximal ventricular contractility \\
$E_{min}$ & mmHg/ml & Minimal ventricular contractility \\
$\tau_{es}$ & s & End systolic time \\
$\tau_{ep}$ & s & End pulse time \\
$Z_{ao}$ & mmHg s/ml & Aortic valve resistance \\
$R_{mv}$ & mmHg s/ml & Mitral valve resistance \\
$R_s$ & mmHg s/ml & Systemic resistance \\
$C_{sa}$ & ml/mmHg & Systemic compliance \\
$C_{sv}$ & ml/mmHg & Venous compliance \\ \bottomrule
\end{tabular}
\begin{tablenotes}
\small
\item The cardiac cycle length $\tau$ is consistently fixed at 1 second, corresponding to a heart rate of 60 BPM.
\end{tablenotes}
\end{threeparttable}
\end{table}

\begin{figure}[ht]
    \centering
    \begin{subfigure}{0.45\linewidth}
        \centering
        \includegraphics[width=\linewidth]{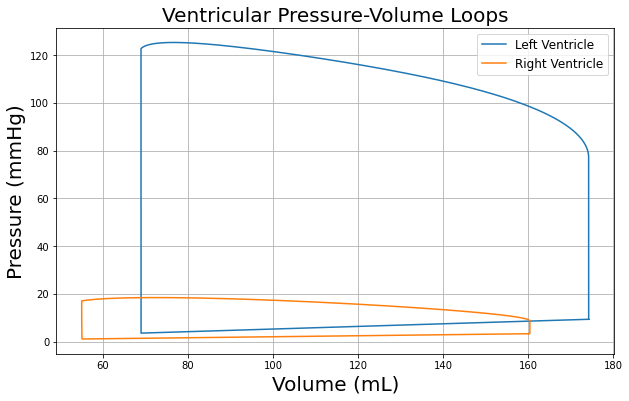}
        \caption{Ventricular PV-Loop}
        \label{fig:2-1}
    \end{subfigure}%
    \hspace{0.05\linewidth}
    \begin{subfigure}{0.45\linewidth}
        \centering
        \includegraphics[width=\linewidth]{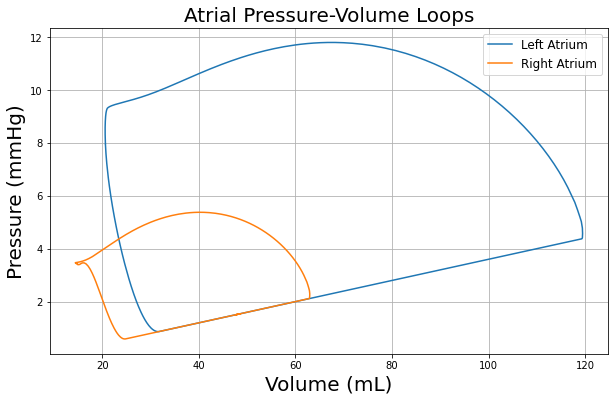}
        \caption{Atrial PV-Loop}
        \label{fig:2-2}
    \end{subfigure}
    \caption{Model 2 outputs of ventricle volumes and atrial volumes, compared with the pressure of each compartment, with sub-interval from 8s to 9s}
    \label{fig:model2-1to2}
\end{figure}

\textbf{Model 2}
\\ \\
After completing the setup for Model 1, we transitioned to Model 2, which shares the same setting as Model 1 but extends to encompass all four chambers, as well as the systemic and pulmonary arteries and veins. The model adheres to the architecture outlined in equations (2), (3), and (4), with similar settings, and expanded compartment subscripts $i$ to all chambers, arteries, and veins (8 in total, cycled through $i$ to $i+7$). However, it employs pressures in terms of elastances and volumes rather than compliances and inflows, which mathematically represents the same concepts as the setting in the first model \cite{colunga2023}:

\begin{equation}
    p_i(t) = E_i(t)V_{s,i}
\end{equation}
\\
Here, $E_{i}(t)$ still represents elastance, but with the inclusion of additional heart chambers, two separate elastance functions are constructed to represent the ventricles and atria, respectively.
\\
Ventricular Elastance Function \( E_v(\tilde{t}) \):
\[
E_v(t) = 
\begin{cases}
    \frac{E_{M,v} - E_{m,v}}{2} \cos\left( \frac{\pi t}{T_{c,v}} \right) + E_{m,v}, & 0 \leq t \leq T_{c,v} \\
    \frac{E_{M,v} - E_{m,v}}{2} \left(1 + \cos\left( \frac{\pi (t - T_{c,v})}{T_{r,v} - T_{c,v}} \right)\right) + E_{m,v}, & T_{c,v} < t \leq T_{r,v} \\
    E_{m,v}, & T_{r,v} < t \leq T
\end{cases}
\]
where:
- \( E_{m,v} \) and \( E_{M,v} \) (mmHg/ml) are the minimal and maximal ventricular elastances.
- \( T_{c,v} \) and \( T_{r,v} \) (s) denote the duration of ventricular contraction and relaxation. \cite{colunga2023}
- \( T \) (s) is the total length of the cardiac cycle, which is fixed at 1 second
\\
Atrial Elastance Function \( E_a(t) \):
\[
E_a(t) = 
\begin{cases}
    \frac{E_{M,a} - E_{m,a}}{2} \left(1 - \cos\left( \frac{\pi (t - T_{r,a})}{T - T_{c,a} + T_{r,a}} \right)\right) + E_{m,a}, & 0 \leq t \leq T_{r,a} \\
    E_{m,a}, & T_{r,a} < t \leq t_{c,a} \\
    \frac{E_{M,a} - E_{m,a}}{2} \left(1 - \cos\left( \frac{\pi (t - t_{c,a})}{T_{c,a} - t_{c,a}} \right)\right) + E_{m,a}, & t_{c,a} < t \leq T_{c,a} \\
    \frac{E_{M,a} - E_{m,a}}{2} \left(1 + \cos\left( \frac{\pi (t - T_{c,a})}{T - T_{c,a} + T_{r,a}} \right)\right) + E_{m,a}, & T_{c,a} < t \leq T
\end{cases}
\]
where:
- \( E_{m,a} \) and \( E_{M,a} \) (mmHg/ml) are the minimal and maximal atrial elastances.
- \( T_{r,a}, t_{c,a}, T_{c,a} \) (s) denote the start of atrial relaxation, the start of atrial contraction, and the point of maximal atrial contraction, respectively. \cite{colunga2023}
- \( T \) (s) is the total length of the cardiac cycle,  which is fixed at 1 second
\\
It is also noteworthy to mention that, as the second model simulates a full cardiovascular cycle, there is no requirement for diode-like behavior similar to the condition outlined in equation (5). 
\\ \\
Based on all the above settings, we can construct and replicate results similar to the original model. This model focuses on analyzing the volumes of all four chambers, systemic and pulmonary arteries, and veins. The remaining variables serve as input parameters, totaling 25 (8 resistances for each output, 4 compliances for each chamber, 8 maximal and minimal elastances for each chamber, and 5 phases of the cardiac cycle). The Table \ref{tab:model2_parameters} contains this information, and the resulting PV-Loops is visualised in Figure \ref{fig:model2-1to2}, which seem successfully replicated the appropriate result from the original model.

\begin{table}[h]
\centering
\begin{threeparttable}
\caption{Model 2 input parameters and their descriptions.}
\label{tab:model2_parameters}
\begin{tabular}{@{}p{3cm}p{3cm}p{8cm}@{}}
\toprule
Parameter & Unit & Description\\ \midrule
\multicolumn{3}{l}{\textbf{Resistances}} \\ \midrule
$R_s$ & mmHg s/ml & Systemic Resistance\\
$R_p$ & mmHg s/ml & Pulmonary Resistance\\
$R_{ava}$ & mmHg s/ml & Aortic Valve Resistance \\
$R_{mva}$ & mmHg s/ml & Mitral Valve Resistance \\
$R_{pva}$ & mmHg s/ml & Pulmonary Valve Resistance \\
$R_{tva}$ & mmHg s/ml & Tricuspid Valve Resistance \\
$R_{pv}$ & mmHg s/ml & Pulmonary Veins Resistance \\
$R_{sv}$ & mmHg s/ml & Systemic Veins Resistance \\ \midrule

\multicolumn{3}{l}{\textbf{Compliances}} \\ \midrule
$C_{sa}$ & ml/mmHg & Compliance of Systemic Arteries \\
$C_{sv}$ & ml/mmHg & Compliance of Systemic Veins \\
$C_{pa}$ & ml/mmHg & Compliance of Pulmonary Arteries \\
$C_{pv}$ & ml/mmHg & Compliance of Pulmonary Veins \\ \midrule

\multicolumn{3}{l}{\textbf{Elastance}} \\ \midrule
$E_{Mra}$ & mmHg/ml & Max Elastance of Right Atrium \\
$E_{mra}$ & mmHg/ml & Min Elastance of Right Atrium \\
$E_{Mla}$ & mmHg/ml & Max Elastance of Left Atrium \\
$E_{mla}$ & mmHg/ml & Min Elastance of Left Atrium \\
$E_{Mrv}$ & mmHg/ml & Max Elastance of Right Ventricle \\
$E_{mrv}$ & mmHg/ml & Min Elastance of Right Ventricle \\
$E_{Mlv}$ & mmHg/ml & Max Elastance of Left Ventricle \\
$E_{mlv}$ & mmHg/ml & Min Elastance of Left Ventricle \\ \midrule

\multicolumn{3}{l}{\textbf{Timing (Phases of cardiac cycle)}} \\ \midrule
$t_{rra}$ & s & Time Right Atrium Relaxation Ends \\
$t_{cra}$ & s & Time Right Atrium Contraction Begins \\
$t_{cra}$ & s & Time Right Atrium Contraction Ends \\
$t_{crv}$ & s & Time Right Ventricle Contraction Begins \\
$t_{rrv}$ & s & Time Right Ventricle Relaxation Begins \\ \bottomrule
\end{tabular}
\begin{tablenotes}
\small
\item The cardiac cycle length $T$ is consistently fixed at 1 second, corresponding to a heart rate of 60 BPM.
\end{tablenotes}
\end{threeparttable}
\end{table}

\subsection{Sensitivity Analysis}

After successfully setting up the models, we will directly apply global sensitivity analysis to all of them to rank each parameter's overall influence to the model output. The Sobol' sensitivity analysis method, advantageous for quantifying the contribution of individual parameters to the total model output variance and capturing interaction effects, will be used to assess parameter importance in the cardiovascular model. By estimating the first-order (direct effect) and total-effect (cumulative influence, including interactions) Sobol indices, we will identify parameters that significantly influence the model outputs, providing a basis for model simplification by fixing unimportant parameters. \cite{nossent2011} This sensitivity analysis will guide the refinement of our model, ensuring a balance between complexity and predictive accuracy.
\\ \\
When conducting global sensitivity analysis (GSA), selecting appropriate parameter ranges is crucial for obtaining meaningful results. The ranges should be wide enough to capture significant variations in model behavior but narrow enough to remain realistic and relevant for the specific application. In other words, it is crucial to limit interactions between parameters to reduce the risk of unrealistic interactions occurring. Therefore, as the first model is simpler and has fewer parameters, applying a wider range (±10\% of the initial value when constructing the model) ensures that the sensitivity analysis captures a broad view of how each parameter affects the model outputs while remaining within a safe boundary. The second model is more complex, with a higher number of parameters and intricate interactions. Thus, the parameter space needs to be managed carefully to avoid unmanageable computational costs and unrealistic results. We will use ±1\% of the initial value as the boundaries for all 25 parameters.
\\ \\
We will maintain all other factors from the model replication unchanged, including the initial parameter values, conditions, time span, and computing accuracy. The SA would not include $S_2$ —typically a second-order sensitivity index that measures the interaction effects between pairs of parameters — due to limited time and computational resources for this project. Instead, the research will only use $S_1$ and $S_T$ indices. $S_1$ is the first-order index that quantifies the direct effect of each parameter on the output variance, while $S_T$ represents the total effect index, capturing both direct and interaction effects. We will analyze the results solely based on the information obtained from these indices; and use different sample size to show their convergence behaviour.
\\ \\
The resulting Sobol indices will be grouped into average Sobol indices based on the mean values for each parameter across all results. The ranking could change if only certain outputs are considered according to clinical demand, as some indices might contribute significantly more to the overall results than others, leading to personalization (depending on the outputs of interest, the list of "important parameters" may differ). However, it's also possible that some parameters remain the most important in all scenarios, indicating their critical influence on model behavior. This consistency could be beneficial for identifying key targets for treatment or intervention, regardless of the specific clinical context. To investigate the potential for ranking variability, we will simulate focusing on only a subset of outputs to demonstrate whether the ranking of importance changes or remains consistent. This simulation will highlight how personalized clinical needs might affect the prioritization of parameters. By analyzing different subsets of outputs, we will be able to determine whether the ranking of parameters shifts significantly, offering valuable insights into which parameters have universally significant effects and which are more context-dependent. Ultimately, this analysis will enhance our understanding of the model's sensitivity landscape, helping to identify robust parameters that consistently affect model performance while also shedding light on parameters that might vary in importance depending on specific clinical scenarios.
\\ \\
We also apply the Morris method sensitivity analysis to Model 1 for a preliminary comparison. The Morris method, a global sensitivity analysis technique, measures the effect of varying one input factor at a time over a predefined number of levels, which we set at $levels = 10$ (the number of levels used to discretize the input factors). The method calculates the "elementary effects" of each input factor, which are the changes in model output resulting from small changes in individual input factors.\cite{yliruka2022}, \cite{balesdent2016} The outputs will include the mean (measuring the overall influence) and standard deviation (indicating non-linearity and interactions) of the elementary effects. We'll normalize these outputs to align with the $S_1$ and $S_T$ indices, enabling a rough comparison with the Sobol method. This approach should ideally yield a similar ranking of parameter importance but with different details.
\\ \\
In the first model, pressure in the left ventricle and the systemic arteries and veins are outputs that cannot be measured using non-invasive methods.
Although there are non-invasive methods of estimating the pressure in the left ventricle when combining the results of brachial blood pressure measurement and magnetic resonance imaging, the left ventricular catheter is the only way to obtain the value of the left ventricle. \cite{arvidsson2023} The pressure in the systemic arteries can be measured using a cuff or stethoscope, and the measured value differs significantly from the actual intra-arterial pressure in particular cases. \cite{meidert2018} In addition, current wearable methods for measurement still face challenges in gaining accurate values. \cite{fortin2021} While the pressure in the systemic veins can be approximated by the combination of 2DE/3DE and acquisition of inferior vena cava volume, challenges for generalisation remain. \cite{szymczyk2020} On the contrary, the volume of left ventricle and inlet volume of the systemic flow can be measured by echocardiography. When employing a transthoracic echocardiogram, electrodes and small sensors are attached to the chest to measure those values. \cite{lang2015} In particular, measuring the inlet volume of the systemic flow, equivalent to cardiac output, by echocardiography is a validated and frequently used method in clinical fields. \cite{zhang2019}
\\ \\
Regarding the second model, more than half of the outputs can be measured by non-invasive methods. For instance, the pulmonary artery's volume can be accurately measured by computer tomography pulmonary angiography or echocardiography. \cite{zhao2023} Echocardiography and cardiovascular magnetic resonance can measure the left and right atria, as well as the left and right ventricles and those measured values are likely to be accurate and reproducible. \cite{lang2015}, \cite{lang2022}, \cite{fent2016} However, we did not identify any non-invasive methods for measuring the blood volume in the systemic arteries, veins, or the pulmonary vein through our literature review.
\\ \\
Table \ref{tab:model_comparison} shows whether each output values in two models can be measured by non-invasive methods. 

\begin{table}[htbp]
    \centering
    \begin{threeparttable}
        \caption{The presence of non-invasive measurements for each model's outputs}
        \label{tab:model_comparison}
        \begin{tabularx}{\textwidth}{@{} >{\hsize=0.8\hsize}X >{\hsize=1.2\hsize}X >{\hsize=0.5\hsize\centering\arraybackslash}X >{\hsize=0.5\hsize\centering\arraybackslash}X @{}}
            \toprule
            \textbf{Output} & \textbf{Description} & \textbf{Model 1} & \textbf{Model 2} \\
            \midrule
            $P_{lv}$ & Pressure in Left Ventricle & No & - \\
            $P_{sa}$ & Systemic Artery Pressure & Yes* & - \\
            $P_{sv}$ & Systemic Vein Pressure & Yes** & - \\
            $Q_{s}$ & Systemic Flow Inlet Volume & Yes & - \\
            $V_{sa}$ & Volume of systemic arteries & - & No \\
            $V_{sv}$ & Volume of systemic veins & - & No \\
            $V_{pa}$ & Volume of pulmonary arteries & - & Yes \\
            $V_{pv}$ & Volume of pulmonary veins & - & No \\
            $V_{ra}$ & Volume of right atrium & - & Yes \\
            $V_{la}$ & Volume of left atrium & - & Yes \\
            $V_{rv}$ & Volume of right ventricle & - & Yes \\
            $V_{lv}$ & Volume of left ventricle & Yes & Yes \\
            \bottomrule
        \end{tabularx}
        \begin{tablenotes}
            \small
            \item[*] Non-invasive methods have limitations in their accuracy
            \item[**] Non-invasive methods need further progress for generalisation
        \end{tablenotes}
    \end{threeparttable}
\end{table}

\subsection{Model Extension}

After successfully replicating the second model, the third model can be easily defined. We will adopt all the settings of the second model, with the only adjustment being the fixing of unimportant parameters, as well as those with high confidence (parameters for which accurate values can be easily obtained through clinical measurements), to reduce the model's complexity. The decision-making will be based on the sensitivity analysis of the second model, selecting parameters with total-effect Sobol indices ($S_T$) below $1/20$ of the average index score, along with the following research on clinical measurements:
\\ \\
In studying the effect of systemic arterial compliance($C_{sa}$) on type 2 diabetes\cite{mohty2012reduced}, systemic arterial compliance was obtained using a non-invasive method by: in line with standard procedure, doppler-echocardiographic measurements were made, which included evaluating the left ventricle's mass, systolic and diastolic function, and end-diastolic and end-systolic diameters. The Dumesnil method, the Simpson method\cite{dumesnil1995new}, and visual estimation were used to evaluate the left ventricular ejection fraction. The left ventricle index (LVi) was obtained by dividing the left ventricle stroke volume in the left ventricle outflow tract by the body surface area. At the same time as the Doppler left ventricle stroke volume measurement in the left ventricle outflow tract, systemic arterial pressure was measured using an arm cuff sphygmomanometer. Systolic and diastolic arterial pressure differences were used to compute brachial pulse pressure (PP), and the ratio of LVi to PP was thought to be an indirect indicator of systemic arterial compliance $C_{sa}$\cite{chemla1998total}.
\\ \\
Although the measurement of minimum elastance of the left ventricle $E_{mlv}$ still requires invasive pressure volume loop analysis in the current routine method \cite{nguyen2020agreement}, a new modified single-beat method has been developed to provide a reference for the measurement of $E_{mlv}$. This method requires five readily measurable parameters that can be obtained via ECG, echo-Doppler cardiography, and noninvasive arm-cuff blood pressure. Systolic $P_s$ and diastolic $P_d$ pressures were measured with an arm cuff sphygmomanometer, stroke volume (SV) with echo-Doppler, ejection fraction (EF) with echocardiography. An estimated normalized ventricular elastance at arterial end-diastole $E_{Nd}$ was calculated using a group-average value that had been modified for each person's contractile and loading effects. Combining the previous parameters, $E_{mlv}$ is calculated as:
\[
E_{mlv} = \frac{P_d - (E_{Nd} \times P_s \times 0.9)}{E_{Nd} \times SV}
\]
In 43 patients with various cardiovascular diseases, the estimates obtained in this manner were compared with invasive measurements. The majority of the absolute differences (75\% to 80\%) between the non-invasive results and the invasive ``target standard'' values were less than 0.6 mm Hg/ml, and an absolute error of 0.6 mm Hg was found sufficient for clinical applicability in most patients in this cohort \cite{chen2001noninvasive}.
\\ \\
Finally, given the state of technology today, intrusive measurements, like right heart catheterization(RHC) \cite{hunter2008pulmonary}, remain the gold standard for obtaining precise data for the majority of physiological parameters in the pulmonary and physical circulation. The above two physiological parameters obtained by non-invasive measurements, after a series of calculations, were able to obtain values of $C_{sa}$ and $E_{mlv}$, which, when compared with the results obtained by invasive methods, were within acceptable limits of error. Therefore, the values of $C_{sa}$ and $E_{mlv}$ obtained by non-invasive measurement methods are highly reliable.

\subsection{Rationality of input parameter values}

Physiological parameters are generally difficult to measure precisely in research, and the majority lack precise reference values. In order to minimize errors caused by inaccuracies in physiological parameters, we refer to a lot of literature to find the normal range of physiological parameters, which can prove our input parameter values reasonable and logical. 
\\ \\
From the references, Ventricular contractility, elastance values in different situations and part of pulmonary resistance parameters are based on parameter selection in a concentrated parameter model for cardiovascular dynamics\cite{korakianitis2006numerical} and a closed-loop lumped parameter model of the left heart and systemic circulation\cite{bjordalsbakke2022parameter}.The choice of values for the resistance of the different vessels and valves from the heart to the whole body and the compliance of the different vascular systems is mainly based on the choice of the relevant parameters in two models: a mathematical model, depicting the short-term regulation of arterial pressure by the carotid baroreceptors under pulsatile conditions.\cite{ursino1998interaction} and a model that represents the
important components of the cardiopulmonary system and their coupled interaction\cite{heldt2002computational}. Some of the parameter values in these reference models are obtained directly from the patient's clinical data or physiological textbooks, while others are calculated by constructing mathematical differential equations based on laws such as Ohm's Law, in combination with physiological parameters that can be obtained directly and easily. Although they are not obtained in the same way and the equations constructed are not exactly the same, the range of parameter values they obtain is basically the same as our input parameter values.
\\ \\ 
The model employs systemic and pulmonary pressure ranges of 70-130 mmHg and 20-30 mmHg, respectively, reflecting typical values observed in healthy human cardiovascular systems. Systemic pressure in the arteries varies significantly during a heartbeat, and this pulsatile nature of blood flow is critical for accurate cardiovascular simulations. In contrast, pulmonary pressures are generally lower due to the lesser resistance and shorter circulation route through the lungs. These distinctions in pressure ranges are essential for creating realistic simulations that can accurately mimic human cardiovascular dynamics.
\\ \\ 
Compliance and resistance values in the model are key to simulating how vessels stretch and recoil, which is vital for understanding cardiovascular dynamics. The compliance of systemic and pulmonary arteries and veins, along with the resistance values, represents the opposition to blood flow through various parts of the circulatory system, impacting blood pressure and flow rates. Additionally, the model incorporates maximum and minimum elastance values for the heart chambers, reflecting their stiffness and relaxation capacity throughout the cardiac cycle. These elastance parameters are crucial for modeling the heart's pumping efficiency and its changes under various cardiac conditions.
\\ \\
The parameter selection in our cardiovascular system model, particularly the resistances and compliances for the systemic and pulmonary circuits, is strongly supported by established practices in cardiovascular modeling. The choice of these specific values is critical for realistically representing the differential dynamics of blood flow in these areas, which vary significantly between systemic and pulmonary circulations due to their distinct physiological functions and mechanical properties.
\\ \\
In this model, resistances such as systemic and pulmonary resistances play crucial roles in defining how blood flows through the heart and lungs. This differentiation is vital because the pulmonary circuit operates under lower pressure and resistance compared to the systemic circuit. The resistance values help in accurately simulating the pressure gradients necessary for effective blood circulation across different vascular beds. Compliance values for the systemic arteries and veins versus pulmonary arteries and veins are chosen to reflect their capacity to stretch and accommodate blood volume changes during the cardiac cycle. Higher compliance in systemic veins, for example, allows them to store more blood at lower pressures, which is characteristic of the systemic venous return to the heart. This aspect is crucial for maintaining venous return and cardiac output under varying physiological conditions.
\\ \\
The inclusion of maximum and minimum elastance values for the chambers of the heart in the models is a critical feature that supports dynamic simulation of cardiac mechanics throughout the cardiac cycle. Elastance, which describes the stiffness of the heart chambers, varies significantly during the cycle, affecting how the chambers fill and eject blood. This is crucial for accurately modeling the heart's mechanical behavior under different physiological and pathological conditions. The heart's mechanical properties change dramatically between the phases of diastole (relaxation) and systole (contraction). Maximum elastance ($E_M$) represents the peak stiffness of the heart chamber during systole, crucial for simulating the powerful ejection of blood. Minimum elastance ($E_m$), on the other hand, reflects the chamber's compliance during diastole, allowing for effective blood filling. These parameters are pivotal in simulating the heart's ability to adapt to varying demands and conditions, such as changes in blood volume or heart rate.
\\ \\
Accurate values for elastance are essential for predicting the heart's response to different stressors and diseases, such as heart failure or valve pathologies. They help in understanding the mechanical aspect of heart function, complementing the electrophysiological models that focus on the propagation of electrical signals. The realistic simulation of these dynamics, supported by robust mathematical definitions of elastance throughout the cardiac cycle, enables the development of targeted therapies and interventions. This modeling approach aligns with current research, which integrates detailed physiological measurements to enhance the predictive power of cardiovascular models.

\newpage
\section{Results}

\subsection{Model 1 Sensitivity Analysis}

\begin{figure}[ht]
    \centering
    \begin{subfigure}{0.32\linewidth}
        \centering
        \includegraphics[width=\linewidth]{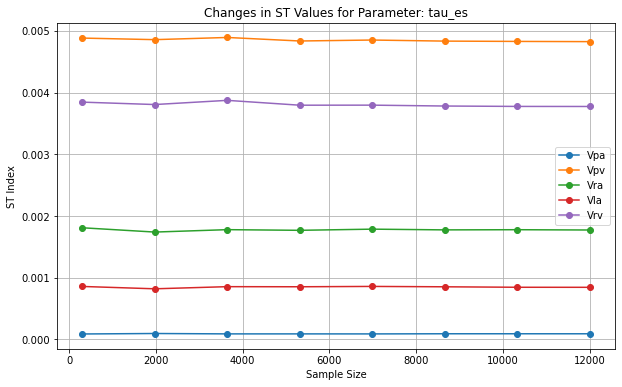}
        \caption{$\tau_{es}$}
        \label{fig:m1-conv-tau_es}
    \end{subfigure}%
    \hspace{0.01\linewidth} 
    \begin{subfigure}{0.32\linewidth}
        \centering
        \includegraphics[width=\linewidth]{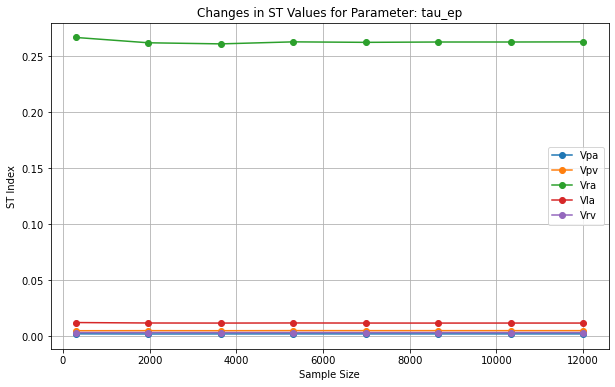}
        \caption{$\tau_{ep}$}
        \label{fig:m1-conv-tau_ep}
    \end{subfigure}%
    \hspace{0.01\linewidth}
    \begin{subfigure}{0.32\linewidth}
        \centering
        \includegraphics[width=\linewidth]{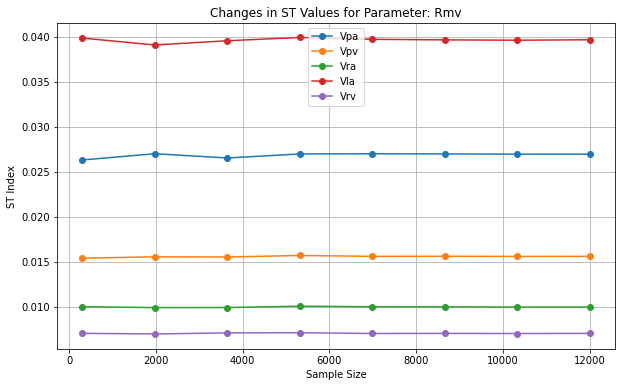}
        \caption{$R_{mv}$}
        \label{fig:m1-conv-R_mv}
    \end{subfigure}
    \hspace{0.01\linewidth}
    \begin{subfigure}{0.32\linewidth}
        \centering
        \includegraphics[width=\linewidth]{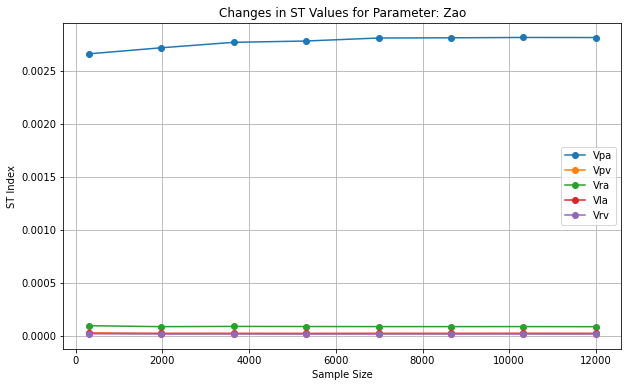}
        \caption{$Z_{ao}$}
        \label{fig:m1-conv-Z_ao}
    \end{subfigure}%
    \hspace{0.01\linewidth} 
    \begin{subfigure}{0.32\linewidth}
        \centering
        \includegraphics[width=\linewidth]{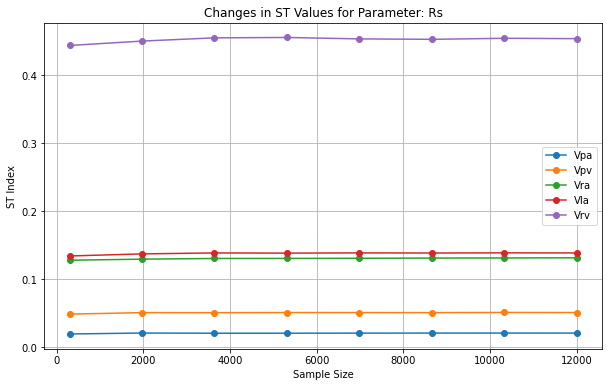}
        \caption{$R_{s}$}
        \label{fig:m1-conv-R_s}
    \end{subfigure}%
    \hspace{0.01\linewidth}
    \begin{subfigure}{0.32\linewidth}
        \centering
        \includegraphics[width=\linewidth]{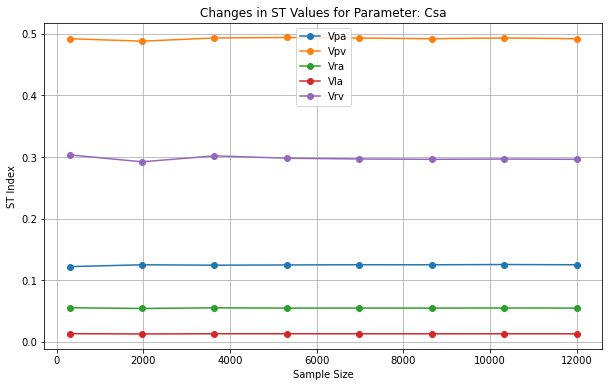}
        \caption{$C_{sa}$}
        \label{fig:m1-conv-C_sa}
    \end{subfigure}
    \hspace{0.01\linewidth}
    \begin{subfigure}{0.32\linewidth}
        \centering
        \includegraphics[width=\linewidth]{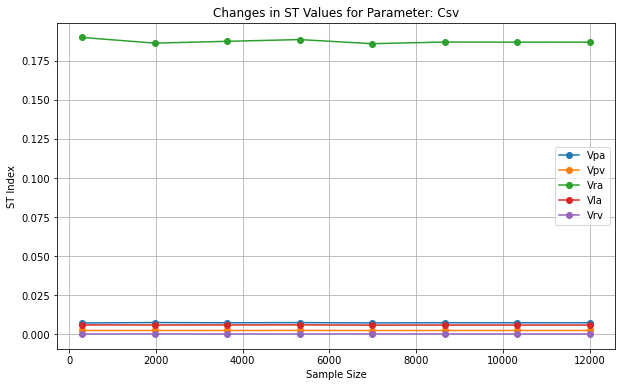}
        \caption{$C_{sv}$}
        \label{fig:m1-conv-C_sv}
    \end{subfigure}%
    \hspace{0.01\linewidth} 
    \begin{subfigure}{0.32\linewidth}
        \centering
        \includegraphics[width=\linewidth]{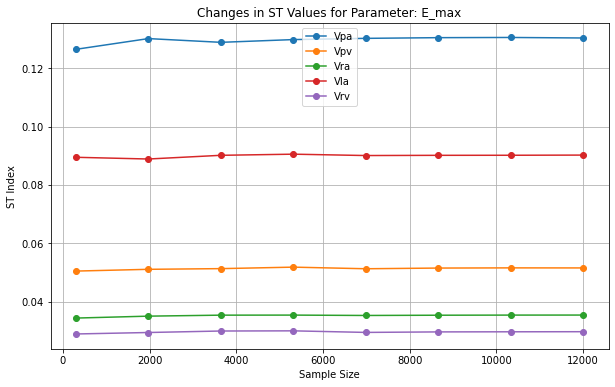}
        \caption{$E_{max}$}
        \label{fig:m1-conv-E_max}
    \end{subfigure}%
    \hspace{0.01\linewidth}
    \begin{subfigure}{0.32\linewidth}
        \centering
        \includegraphics[width=\linewidth]{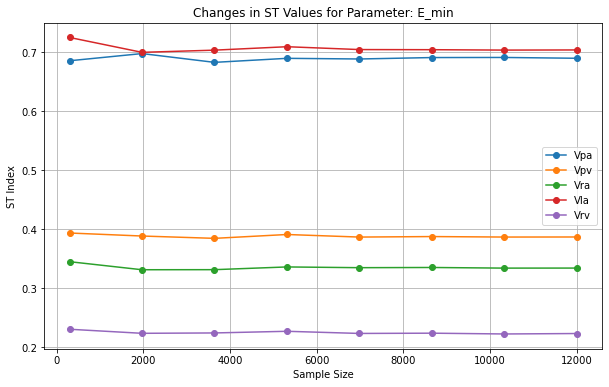}
        \caption{$E_{min}$}
        \label{fig:m1-conv-E_min}
    \end{subfigure}
    \hspace{0.01\linewidth}
    \caption{The convergence pattern of each parameter across different sample sizes of [300, 500, 1000, 1500, 2500, 4000, 6000, 12000]. The x-axis is the number of sample sizes and the y-axis is the value of the Sobol indices under each parameter size}
    \label{fig:model1-cov}
\end{figure}

\begin{figure}[ht]
    \centering
    \begin{subfigure}{0.45\linewidth}
        \centering
        \includegraphics[width=\linewidth]{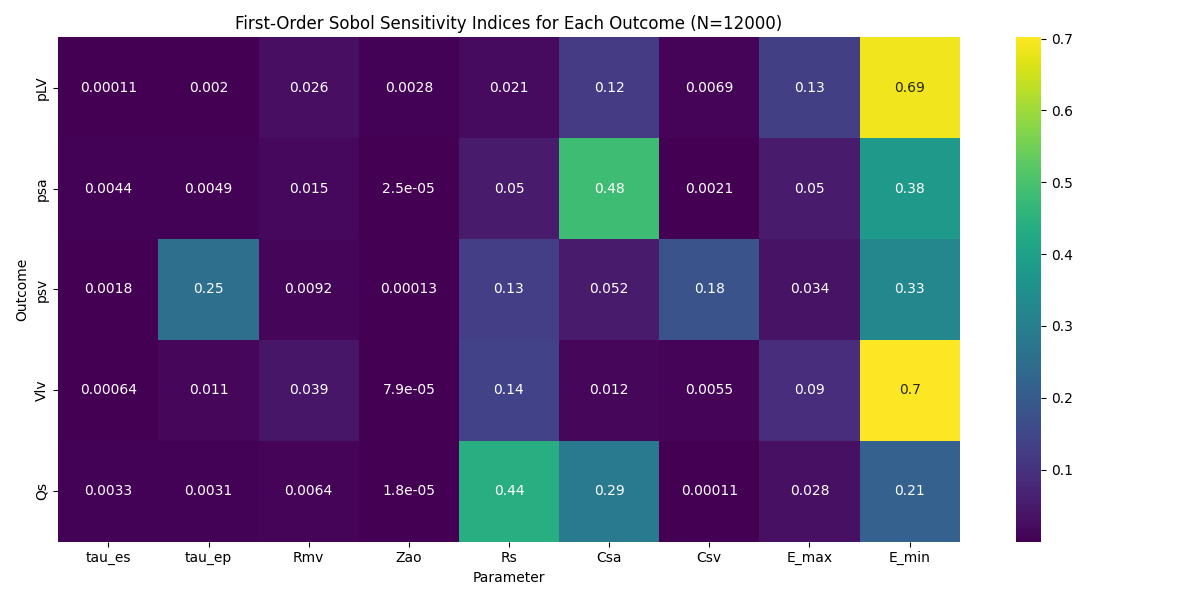}
        \caption{$S_1$ values as heatmap}
        \label{fig:m1-S1-heatmap}
    \end{subfigure}%
    \hspace{0.05\linewidth} 
    \begin{subfigure}{0.45\linewidth}
        \centering
        \includegraphics[width=\linewidth]{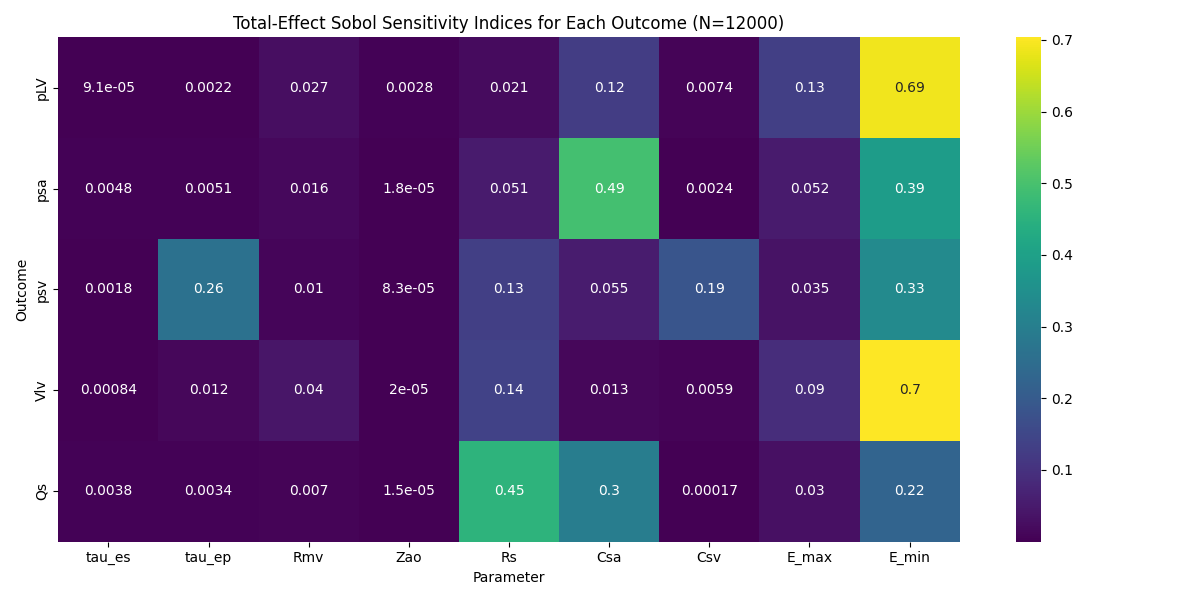}
        \caption{$S_T$ values as heatmap}
        \label{fig:m1-ST-heatmap}
    \end{subfigure}%
    \hspace{0.05\linewidth}
    \begin{subfigure}{0.45\linewidth}
        \centering
        \includegraphics[width=\linewidth]{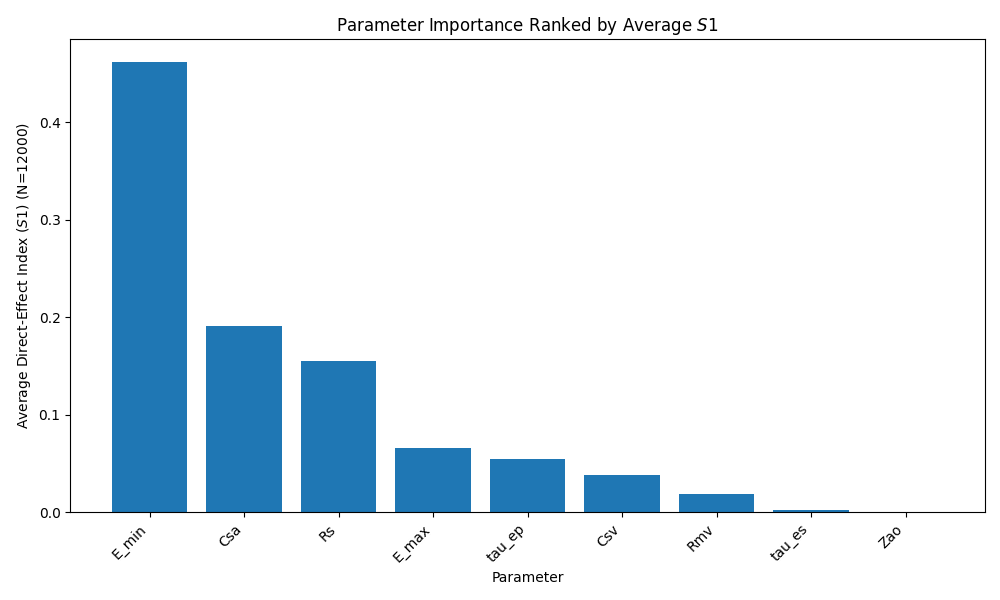}
        \caption{Importance Ranking of $S_1$}
        \label{fig:m1-S1-rank}
    \end{subfigure}
    \hspace{0.05\linewidth}
    \begin{subfigure}{0.45\linewidth}
        \centering
        \includegraphics[width=\linewidth]{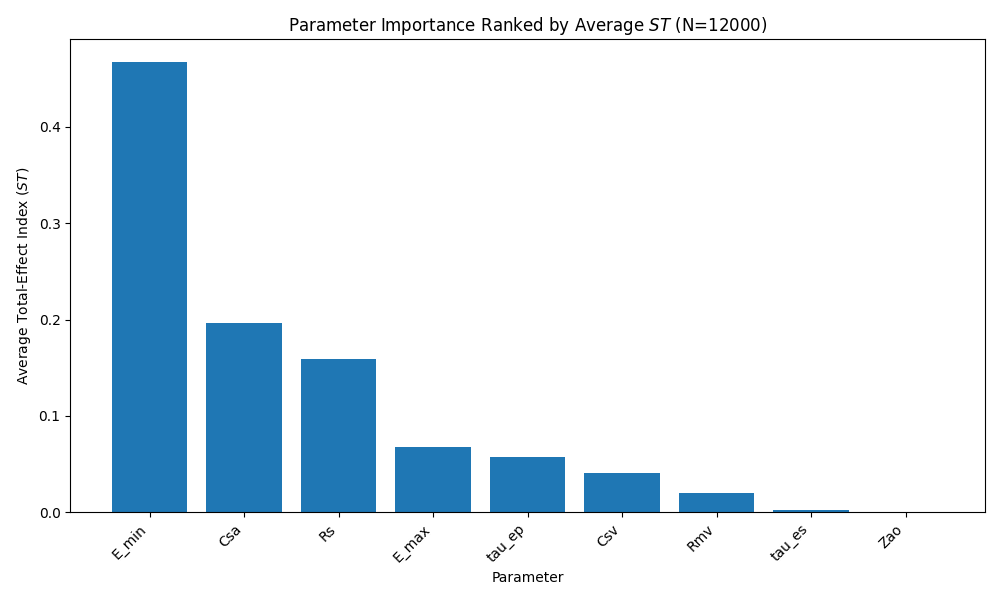}
        \caption{Importance Ranking of $S_T$}
        \label{fig:m1-ST-rank}
    \end{subfigure}
    \hspace{0.05\linewidth}
    \caption{Sobol Method Result of Model 1: figure(a) is the first-order sensitivity indices of Model 1 and figure(b) is the total-effect sensitivity indices. Figure(c) is the rank of parameter importance of figure a) according to the mean value of each column, and figure(d) is the rank of parameter importance of figure(b)}
    \label{fig:model1-S1to4}
\end{figure}

\textbf{Sobol}\\
\\
We have firstly examined the convergence of the results by independently analyzing each parameter’s behavior through their respective Sobol indices and how they contribute to each output. The stability of the Sobol indices for most parameters, even with a relatively low number of samples, indicates robust model performance. Convergence with minimal error typically occurs between 2,500 to 6,000 samples. This observation in figure \ref{fig:model1-cov} suggests that our model achieves reliable sensitivity outcomes before further sample sizes. 
\\
\\
The result for Sobol Method is contained in figure \ref{fig:model1-S1to4}. The similarity between the $S_1$ and $S_T$ indices after the sensitivity analysis (SA) converges is surprisingly close, displaying only slight differences in the Sobol indices. Before convergence, this similarity was not observed. The $S_1$ index quantifies the proportion of the output variance attributable solely to the main effects of individual input parameters, while $S_T$ considers both the main effects and the interactions of each parameter with others. This close alignment suggests that interactions between parameters minimally contribute to the variability of the model outputs, indicating that the model outcomes are primarily influenced by the direct effects of individual parameters. Such a scenario is typical in other linear models where the contributions of parameters are independent of each other.
\\ 
\\
\\
This observation could stem from the choice of initial parameter values or the limited range used for the sensitivity analysis, possibly restricting the detection of more complex interactions. These results suggest that expanding the range or reassessing the initial settings might reveal additional dynamics. Further exploration into these aspects could provide deeper insights into the underlying mechanisms driving the model, potentially uncovering significant interactions that were previously undetected. \\
\\
The next step is to analyse the result of the parameter importance ranking based on figure c) and d): $E_{min}$ (Minimal ventricular contractility) emerges as the most influential parameter, having the highest $S_T$ value, which indicates its significant impact on the variance of the model's outputs. Following behind are $C_{sa}$ (Compliance of Systemic Arteries) and $R_s$ (Systemic Resistance), with $C_{sa}$ being the second most influential, though its $S_T$ value is significantly lower than $E_{min}$, indicating a lesser but still notable impact on the model's output. $R_s$ follows closely behind $C_{sa}$, suggesting its crucial role in the system. $E_{max}$ (Maximal ventricular contractility) and $\tau_{ep}$ (End pulse time) show moderate influence, which indicates that while they affect the model, their impact is not as strong as the top parameters. $C_{sv}$ (Compliance of Systemic Veins), $R_{mv}$ (Mitral Valve Resistance), and $\tau_{es}$ (End systolic time) have relatively lower $S_T$ values, suggesting that their contribution to the model's output variance is less critical under the tested conditions. Lastly, $Z_{ao}$ has the lowest $S_T$ index, indicating minimal impact on the model's outputs relative to other parameters. These results suggests that while some parameters predominantly drive the model's output, others have minimal effects. These information reveals several critical implications for the model. The high importance of $E_{min}$ underscores the necessity of ensuring accurate estimation and modeling of this parameter to maintain the reliability and accuracy of the model. Parameters with very low $S_T$ indices, such as $Z_{ao}$, might be candidates for simplification in the model, potentially reducing computational complexity without significantly affecting model accuracy. Meanwhile, parameters that show moderate to high $S_T$ values should be the focus of further investigation to comprehensively understand their roles and interactions within the model.
\newpage

\begin{figure}[ht]
    \centering
    \begin{subfigure}{0.45\linewidth}
        \centering
        \includegraphics[width=\linewidth]{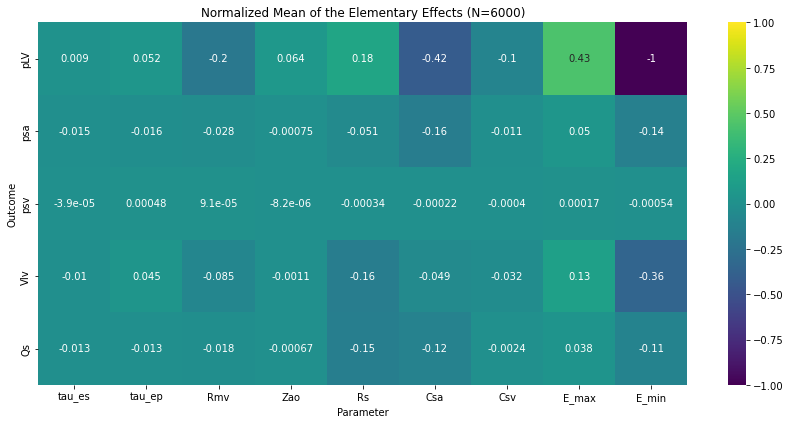}
        \caption{Mean elementary effects}
        \label{fig:m1-morris-mean}
    \end{subfigure}%
    \hspace{0.05\linewidth} 
    \begin{subfigure}{0.45\linewidth}
        \centering
        \includegraphics[width=\linewidth]{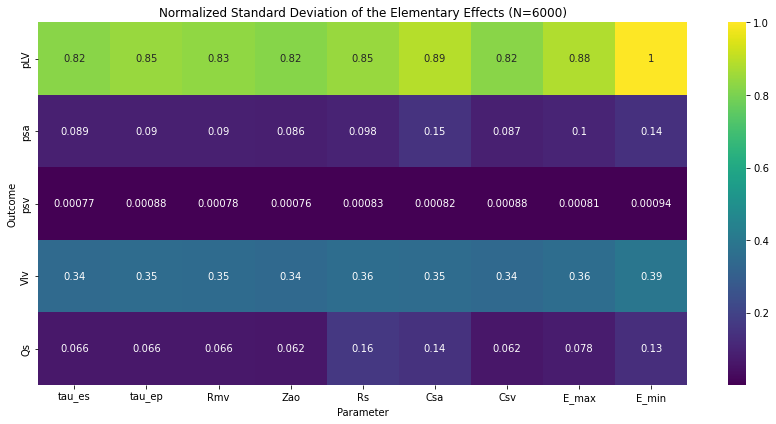}
        \caption{Standard deviation of elementary effects}
        \label{fig:m1-morris-std}
    \end{subfigure}%
    \hspace{0.05\linewidth}
    \begin{subfigure}{0.8\linewidth}
        \centering
        \includegraphics[width=\linewidth]{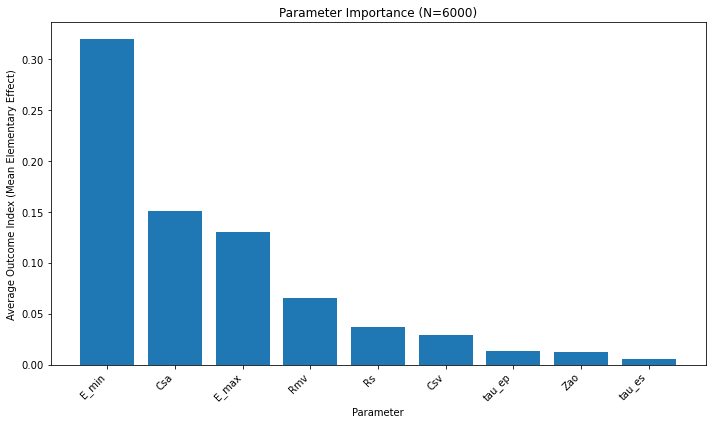}
        \caption{Importance Ranking}
        \label{fig:m1-morris-rank}
    \end{subfigure}
    \hspace{0.05\linewidth}
    \caption{Morris Method Result for Model 1: figure(a) is the mean elementary effect; the values are normalized into range of $[-1,1]$ to show their relative effect as well as directions. Figure(b) is the standard deviation of the elementary effects, normalized into range of $[0,1]$. Figure(c) is the rank of parameter importance, based on the absolute average value of each parameter from figure(a)}
    \label{fig:model1-M1to3}
\end{figure}

\vspace{\baselineskip}\textbf{Morris}\\
\\
Since the first model is relatively straightforward and involves fewer parameters, we will compare its results with those from the Morris Method instead of applying simplification. Testing has shown that the time taken for different sample numbers in both methods is strictly linear; this means that 12,000 samples will take exactly twice as long as 6,000 samples. Given that the model converges quickly, we have opted to use 6,000 samples for the Morris method for comparison. The results are presented in Figure \ref{fig:model1-M1to3}, and it is notably evident that the Morris method consumes significantly fewer computational resources compared to the Sobol method — it required approximately 4.74 times more time for Sobol than for Morris. 
\\ 
\\
The first figure illustrates the normalized mean of the elementary effects for each parameter across different outputs, where elementary effects measure the impact of slight changes in each parameter on the model output. A higher mean value indicates a stronger influence of the parameter on the model’s output, with color coding from dark green to purple (and yellow for extreme values) depicting a range from low to high impacts. Parameters such as $E_{min}$, $C_{sa}$, and $E_{max}$ show significant variability across outputs, highlighting their substantial effects on model behavior, with some values indicating inverse relationships between parameter changes and outputs. The second figure shows the normalized standard deviation of the elementary effects, quantifying the dispersion of effects each parameter has on the outputs. Higher standard deviations, as seen with parameters like $E_{min}$ and $E_{max}$, suggest that these parameters introduce significant variability into the model’s output, especially under different conditions or parameter combinations within the model.
\\ \\
The third figure provides an analysis of parameter importance via the Morris Method, plotting the average outcome index (mean elementary effect) for each parameter and reflecting their overall influence on model outputs. Both the Morris and Sobol methods consistently identify $E_{min}$ as the most influential parameter, underscoring its pivotal role across different analytical approaches. Similarly, $C_{sa}$ and $E_{max}$ are highlighted as significant in both analyses, confirming their critical impact on the model's behavior. Notably, parameters such as $R_{mv}$ and $R_s$, while moderately influential in the Sobol analysis, appear much less or more impactful in the Morris results, suggesting that their perceived importance may vary significantly depending on the method employed. This variance highlights that certain parameters might be method-dependent, potentially influencing decision-making regarding model adjustments or further explorations. The consistency in the ranking of parameters like $E_{min}$, $C_{sa}$, and $E_{max}$ across both methods establishes them as stably important, whereas others' importance could fluctuate based on the sensitivity analysis technique used.

\subsubsection{Validation 1}

The measurements discussed above are based on conditions where all possible information is gathered. However, as mentioned in Section 3.4, it is not always feasible to measure all outputs from sensitivity analysis due to various factors, including patient risks and technical limitations. Consequently, the model adjustments will be justified based on the information provided in Table \ref{tab:model_comparison} in Section 3.4 regarding non-invasive outputs. Specifically, we will exclude the output of $p_{LV}$ (Pressure in Left Ventricle) and $p_{sa}$ (Systemic Artery Pressure) to re-rank the importance of parameters. The exclusion is warranted because $p_{LV}$ requires invasive measurements in all cases, and $p_{sa}$ faces limitations in the accuracy of potential non-invasive measurements. The conclusion is included in figure \ref{fig:model1-S1A} below.
\\
\begin{figure}[ht]
    \includegraphics[width=\linewidth]{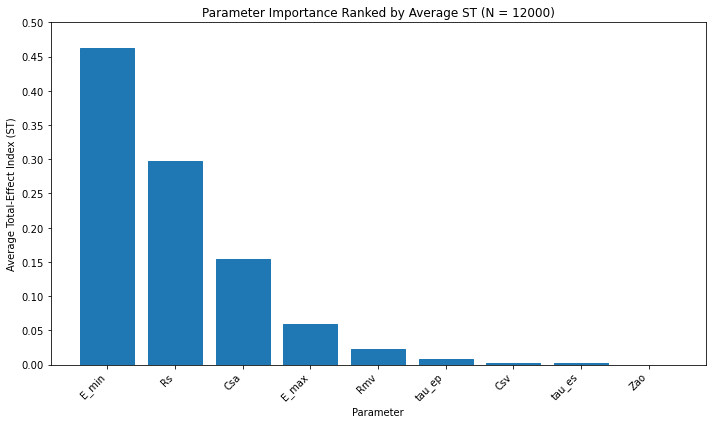}
    \caption{Rearranged parameter importance without influence from $p_{LV}$, $p_{sa}$, and $p_{sv}$ of Model 1 SA result for $S_T$ of Sobol method with sample size = 12000}
    \label{fig:model1-S1A}
\end{figure}
\\
The comparative analysis of the latest parameter importance graph against previous Sobol results highlights consistent and variable influences across different settings. In both datasets, $E_{min}$ (Minimum Elastance) emerges as the most impactful parameter, consistently demonstrating the highest Average Total-Effect Index $S_T$, underscoring its crucial role in the model's dynamics. Parameters such as $R_s$ (Systemic Resistance) and $C_{sa}$ (Systemic Artery Compliance) maintain strong influences, although their rankings vary slightly, indicating sensitivity to model conditions or analysis methods. Notably, $E_{max}$ (Maximum Elastance) and $C_{sv}$ (Systemic Vein Compliance) show significant shifts in importance, suggesting a context-dependent influence on model outputs. Both analyses agreed on the lower influence of parameters like $\tau_{es}$ (End Systolic Time) and $Z_{ao}$, which consistently rank lower, indicating their minimal impact under tested conditions. This consistent and variable parameter ranking is crucial for model refinement and suggests further investigation into parameters whose influence varies across different scenarios or conditions.

\subsection{Model 2 Sensitivity Analysis}

\begin{figure}[ht]
    \centering
    \begin{subfigure}{0.45\linewidth}
        \centering
        \includegraphics[width=\linewidth]{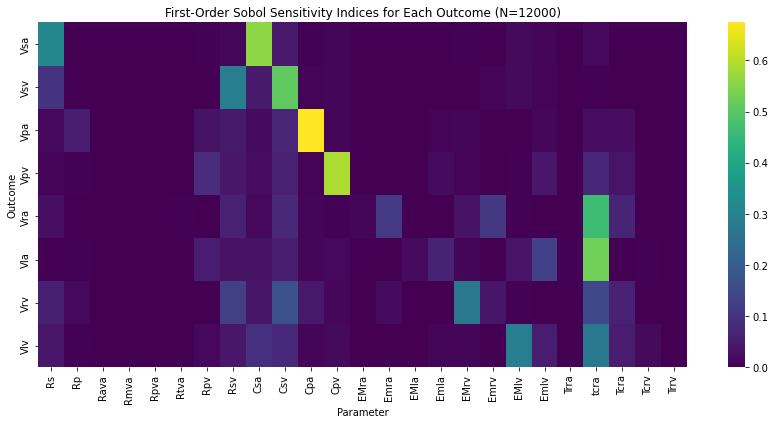}
        \caption{$S_1$ values as heatmap}
        \label{fig:m2-S1-heatmap}
    \end{subfigure}%
    \hspace{0.05\linewidth} 
    \begin{subfigure}{0.45\linewidth}
        \centering
        \includegraphics[width=\linewidth]{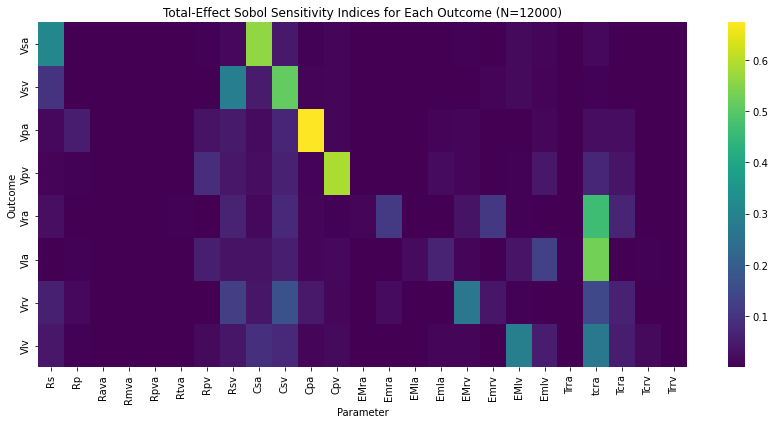}
        \caption{$S_T$ values as heatmap}
        \label{fig:m2-ST-heatmap}
    \end{subfigure}%
    \hspace{0.05\linewidth}
    \begin{subfigure}{0.45\linewidth}
        \centering
        \includegraphics[width=\linewidth]{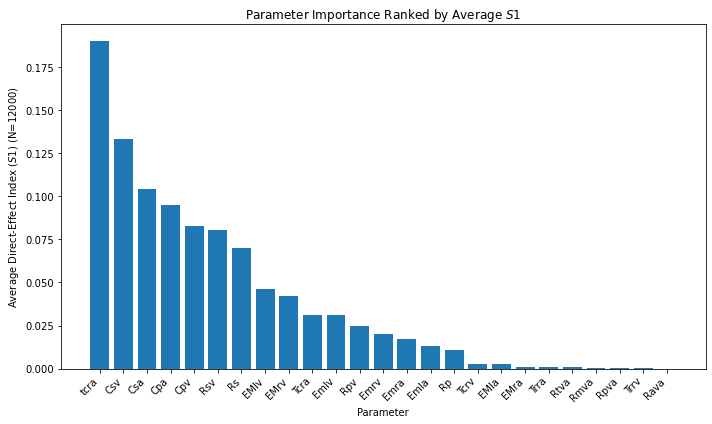}
        \caption{Importance Ranking of $S_1$}
        \label{fig:m2-S1-rank}
    \end{subfigure}
    \hspace{0.05\linewidth}
    \begin{subfigure}{0.45\linewidth}
        \centering
        \includegraphics[width=\linewidth]{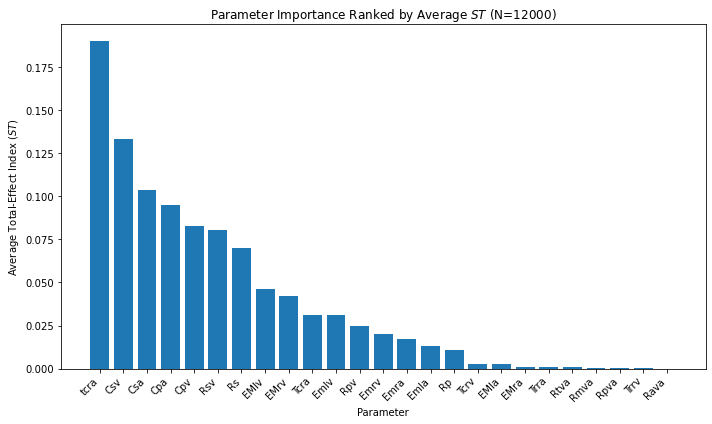}
        \caption{Importance Ranking of $S_T$}
        \label{fig:m2-ST-rank}
    \end{subfigure}
    \hspace{0.05\linewidth}
    \caption{Sobol Method Result of Model 2: figure(a) is the first-order sensitivity indices of Model 2 and figure(b) is the total-effect sensitivity indices. Figure(c) is the rank of parameter importance of figure a) according to the mean value of each column, and Figure(d) is the rank of parameter importance of figure(b)}
    \label{fig:model2-S1to4}
\end{figure}

Like the sensitivity analysis results from Model 1, Model 2 adopts the same structure but with a significantly larger number of parameters to rank. The results visualised in Figure \ref{fig:model2-S1to4} also encounters a similar issue as detailed previously: the similarity between the $S_1$ and $S_T$ indices after the SA converges is notably close, reflecting minimal differences in the Sobol indices. This convergence suggests that the model outcomes are primarily influenced by the direct effects of individual parameters rather than their interactions. Notably, despite the increased complexity of Model 2, it uses even narrower SA boundaries, which may limit the detection of more nuanced interactions. 
\\
\\
The parameter $t_{cra}$ (Time Right Atrium Contraction Begin) emerges as the most impactful, reflecting its significant direct effect on the model's outputs. This timing parameter is crucial for understanding atrial dynamics, particularly how atrial contraction influences ventricular filling. Most of Compliances such as $C_s$ (Compliance of Systemic Arteries) and $C_{sv}$ (Compliance of Systemic Veins) also rank high, underscoring the importance of vascular compliance in modulating systemic hemodynamic responses. Meanwhile, resistances like $R_s$ (Systemic Resistance) and $R_p$ (Pulmonary Resistance) are in general slightly less important than $t_{cra}$ and compliance, affecting overall cardiovascular function. Elastance parameters, especially $E_{max}$ and $E_{min}$ for various heart chambers, including the left ventricle $E_{Mlv}$ and $E_{mlv}$, show moderate importance, which indicates their role in determining the heart's contractile and relaxation capabilities essential for cardiac output. Other parameters, such as $R_{mva}$ (Mitral Valve Resistance) and $\tau_{es}$ (End Systolic Time), though influential, have a comparatively lower direct effect on the model outcomes. The least influential parameters, like $\tau_{ep}$ (End Pulse Time) and $R_{ava}$ (Aortic Valve Resistance), suggest that changes in these areas, within the tested conditions, have minimal direct impact on overall model behavior.

\subsubsection{Validation 2}

Similarly, examine the avoidance of invasive measurements in the context of the four-chamber model of full cardiovascular circulation prompts a revisitation of Table \ref{tab:model_comparison} in Section 3.4. Herein, the parameters $V_{sa}$, $V_{sv}$, and $V_{lv}$ emerge as outputs lacking non-invasive measurement methodologies. Consequently, their exclusion from analysis is imperative to assess the model's stability across diverse scenarios. By omitting these variables, we can test the model's robustness under varied conditions, highlighting on its reliability and applicability in non-invasive measurement contexts. The following figure \ref{fig:model2-S1A} would be the rearranged parameter importance based on the rest of outputs from model 2 SA.
\\
\begin{figure}[ht]
    \includegraphics[width=\linewidth]{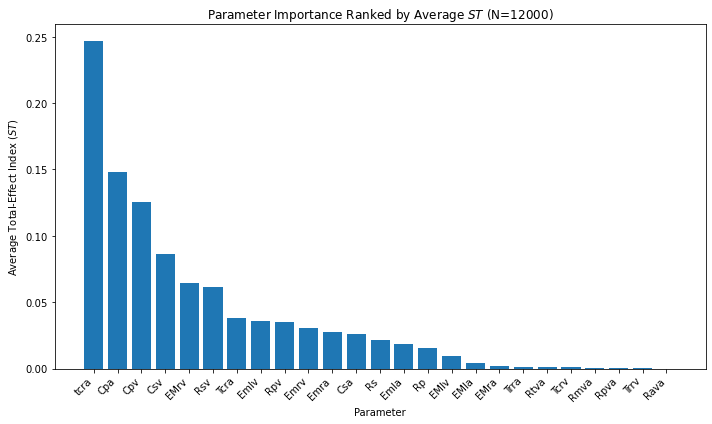}
    \caption{Rearranged parameter importance without influence from $V_{sa}$, $V_{sv}$, and $V_{lv}$ of Model 2 SA result for $S_T$ of Sobol method with sample size = 12000}
    \label{fig:model2-S1A}
\end{figure}
\\
The comparative analysis of two figures ranking parameter importance based on the Average Total-Effect Index $S_T$ from sensitivity analyses reveals both stability and variability in parameter impacts. Notably, $t_{cra}$ (Time Right Atrium Contraction Ends) emerges as the highest-ranking parameter in both figures, demonstrating its substantial and consistent influence on the model's outputs. This suggests a crucial role for the timing of atrial contraction in the cardiovascular dynamics modeled. On the other hand, $C_{sa}$ (Compliance of Systemic Arteries) experiences a significant drop in importance, moving from third place in the first figure to twelfth in the second. This shift indicates a potentially variable perception of how systemic arterial compliance impacts the model, depending on the specifics of the sensitivity analysis or the scenarios tested. Other parameters such as $C_{pv}$ and $C_{sv}$ (Compliance of Pulmonary and Systemic Veins respectively) maintain high ranks, underscoring their continued perceived importance in influencing model dynamics. Resistance parameters like $R_s$ (Systemic Resistance) and $R_p$ (Pulmonary Resistance) consistently appear in the middle to lower end of the importance spectrum, reflecting a less dominant yet notable influence compared to compliance and timing parameters. The low rank parameters remains much less influential as expected, as they would not gain increases in sobol indices. The variation observed, particularly with $C_{sa}$, highlights the nuanced nature of parameter interactions and the potential impact of different modeling assumptions or analysis setups on the perceived importance of specific model parameters. This analysis underscores the need for careful parameter evaluation and validation in sensitivity analyses to ensure robust and reliable model insights, but also illustrated the relative stability of the model

\subsection{Model3 Sensitivity Analysis}

Initially, we will discard the less significant parameters, those with Sobol indices below 5\% of the mean indices across all parameters. According to the results detailed in section 5.2, these parameters include $t_{crv}$, $E_{Mla}$, $E_{Mra}$, $t_{rra}$, $R_{tva}$, $R_{mva}$, $R_{pva}$, $t_{rrv}$, and $R_{ava}$. Subsequently, we will remove parameters that are either readily measurable through non-invasive techniques or are highly reliable and well-established, specifically $C_{sa}$ and $E_{mlv}$ as discussed in section 4.4.
\\
\begin{figure}[ht]
    \centering
    \begin{subfigure}{0.45\linewidth}
        \centering
        \includegraphics[width=\linewidth]{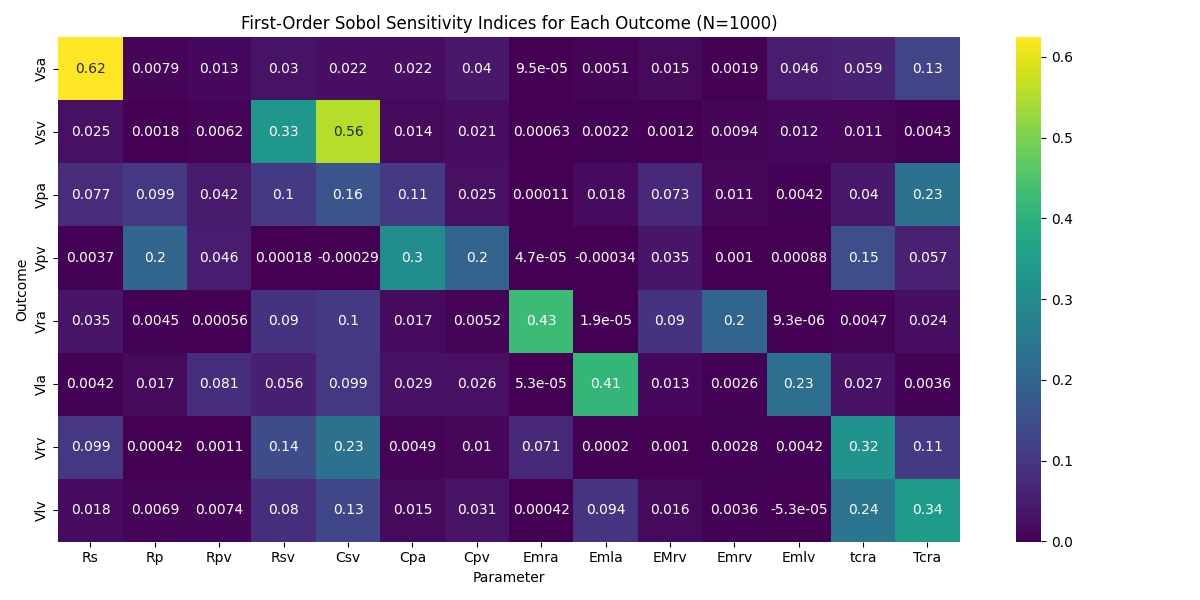}
        \caption{$S_1$ values as heatmap}
        \label{fig:m3-S1-heatmap}
    \end{subfigure}%
    \hspace{0.05\linewidth} 
    \begin{subfigure}{0.45\linewidth}
        \centering
        \includegraphics[width=\linewidth]{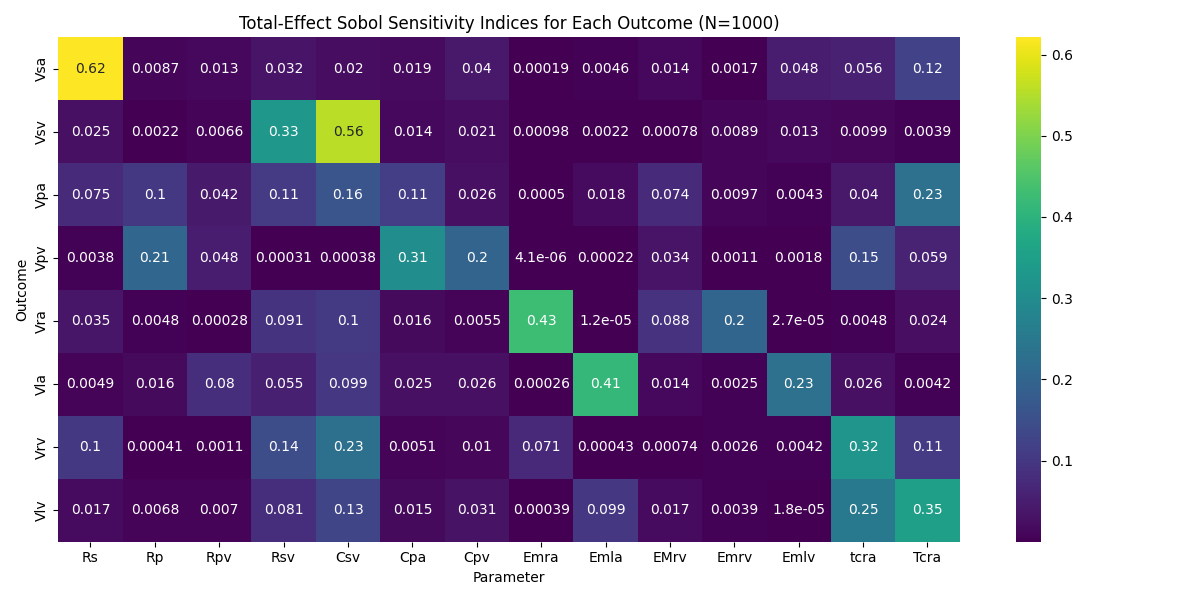}
        \caption{$S_T$ values as heatmap}
        \label{fig:m3-ST-heatmap}
    \end{subfigure}%
    \hspace{0.05\linewidth}
    \begin{subfigure}{0.8\linewidth}
        \centering
        \includegraphics[width=\linewidth]{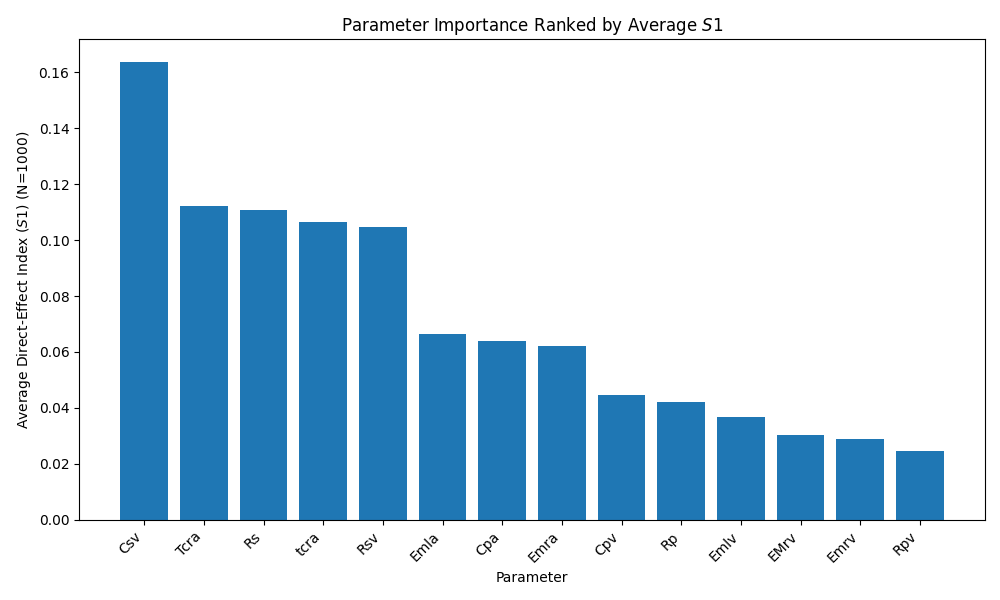}
        \caption{Importance Ranking of $S_1$}
        \label{fig:m3-S1-rank}
    \end{subfigure}
    \hspace{0.05\linewidth}
    \hspace{0.05\linewidth}
    \caption{Sobol Method Result of Model 3: figure a) is the first-order sensitivity indices of Model 3 and figure b) is the total-effect sensitivity indices. Figure c) is the rank of parameter importance of figure a) according to the mean value of each column}
    \label{fig:model3-S1to3}
\end{figure}
\\
As a result of limited time available for the project, we will conduct a simulation with a sample size of 1,000 instead of the originally intended 12,000 samples. Although this approach will lead to a less precise simulation, the model is anticipated to stabilise rapidly even with a smaller number of samples, as suggested by preliminary testing with a limited sample size and previous analyses showing rapid convergence. Hence, we are proceeding with the presumption that the outcomes obtained from employing 1,000 samples will closely approximate those obtained from using 12,000 samples. All subsequent analyses and discussions will be conducted under the assumption, acknowledging that it may not accurately represent the actual condition. \\
\\
The result is included in figure \ref{fig:model3-S1to3}. Since Model 3 experiences the same issue as Model 2, with the $S_1$ and $S_T$ values being nearly identical (This is obvious in the heatmaps) due to shared settings, we will present only one figure for parameter importance ranking to better represent the data. Comparing the result of model 2 SA, a notable changes in parameter importance and the overall distribution of rankings is observed. The new model has streamlined the focus onto fewer variables, leading to a clearer view of which parameters significantly drive the model outputs without the previously included parameters' influence. In this simplified model, $C_{sv}$(Compliance of Systemic Veins) emerges as the most influential parameter from the second place, followed by $T_{cra}$ (Time Right Atrium Contraction Ends) which was far less influential in model 2, $R_s$ (Systemic Resistance), $t_{crv}$ (Time Right Ventricle Contraction Begins), and $R_{sv}$ (Pulmonary Veins Resistance). Pulmonary compliances namely $C_{pa}$ and $C_{pv}$ received a significant drop in importance, while the other parameters remains stable. Parameters like $C_{sv}$ remain highly influential in both plots, stating the similarity between two models even with the elimination of the 11 parameters. The results appear to be relatively stable in half of the high-ranking and most of the low-ranking parameters between the two models, particularly which suggests that the model retains its main behaviour characteristics even with fewer parameters. However, it is also important to highlight the significant shifts in $T_{cra}$, $C_{pa}$, and $C_{pv}$, which indicate unexpected fluctuations. This suggests that even parameters deemed unimportant or reliable by the sensitivity analysis might still impact the model's overall performance in certain contexts, particularly when specific parameters are implicitly interconnected.

\section{Discussion and Conclusions}

\subsection{Models and Results}

As outlined in the Introduction and Methodology sections, our study primarily utilised Sobol sensitivity analysis methods, applied across three distinct cardiovascular models. Model 1 employs a simpler system focusing on key cardiovascular parameters, while Model 2 expands on this by incorporating a four-chamber heart representation, providing a more complex and detailed simulation environment. The Sobol method applied to Model 1 allowed for the quantification of the contribution of each parameter to the output variance, capturing both main effects and interactions. The analysis demonstrated that certain parameters, notably minimal ventricular contractility ($E_{min}$) and systemic arterial compliance ($C_{sa}$), had a substantial influence on the model's outputs. The Sobol indices suggested a predominance of first-order effects with minimal interaction between parameters, indicating that the model's behaviour under the initial value of parameters and their range concerned could largely be predicted from individual parameter changes.\\
\\
Conversely, the Morris method has been applied to Model 1 for comparison. It evaluates the elementary effects of parameter variations one at a time, providing a faster yet less detailed overview compared to the nature of Sobol indices. This method identified similar key parameters as the Sobol method but required significantly less computational time. For instance, the Morris method took approximately one-fourth to one-fifth the time of the Sobol method to complete, offering a much more computationally efficient alternative for preliminary sensitivity analysis. The comparative analysis between the Sobol and Morris methods in Model 1 underscores the trade-off between computational demand and analytical depth. While the Sobol method offers detailed insights into parameter interactions, the Morris method, with its reduced computational load, provides a quick snapshot of parameter sensitivities, making it ideal for initial assessments.\\
\\
Model 2's detailed representation of the cardiovascular system required a more complex sensitivity analysis to effectively manage its broad range of interactions. The Sobol method, used in this analysis, underscored the consistently significant role of certain key parameters, notably $t_{cra}$. The results from the sensitivity analysis of Model 2 showed that the model could accurately simulate detailed cardiovascular behaviours and adapt to physiological variations. This level of detail in the simulations highlights the model's effectiveness in reflecting real-world clinical conditions, thereby enhancing its usefulness in clinical research and the development of personalised treatments for individual patients. These insights demonstrate the value of advanced computational models in medical research, where a deep understanding of the interactions within cardiovascular systems can lead to better decisions in patient care and treatment planning. The findings align with the goals outlined in the Introduction, where we aimed to study advanced modelling techniques to improve the predictive accuracy and clinical relevance of cardiovascular studies.\\
\\
For Model 3, our analysis explored the possibility of simplifying the model by excluding parameters identified as both unimportant and confidently measured through the sensitivity analysis. While this approach holds promise for reducing model complexity and computational demands, our findings indicate that a deeper understanding of the interactions between parameters is crucial. Simplification based solely on Sobol indices of individual importance may overlook how these parameters interact in complex ways that significantly influence the model's overall performance. This underscores the need for careful tuning and understanding of the model when adjustments are made. Precise calibration of the parameters and thorough testing of the model's response to these changes are essential to ensure that simplifications do not compromise the accuracy or the predictive capabilities of the model, especially in replicating complex cardiovascular behaviours. This nuanced approach will enhance the model’s utility in clinical research, providing a balanced perspective between simplicity and detailed dynamism in cardiovascular simulations.

\subsection{Validation according to non-invasive Measurements (5.1.1 and 5.2.1)}

In Model 1, the adjustments involved re-evaluating the importance of parameters when invasive measures, like the pressure in the left ventricle $p_{LV}$ and systemic artery pressure $p_{sa}$, are not feasible. By excluding them from the analysis, the revised parameter importance rankings highlighted the persistent significance of Minimum Elastance $E_{min}$, systemic resistance $R_s$, and systemic artery compliance $C_{sa}$. This exclusion directly affects clinical adaptability, as it tailors the model to use parameters that can be assessed via non-invasive techniques, thus broadening the model's utility in standard clinical practices where minimising patient risk is paramount.\\
\\
Similarly, Model 2 adjustments considered outputs lacking non-invasive measurement methodologies, namely the volumes of systemic arteries ($V_{sa}$) and veins ($V_{sv}$), and the left ventricle ($V_{lv}$). Excluding these parameters from sensitivity analysis necessitated a re-assessment of other parameters' impacts. The revised rankings demonstrated that parameters like right atrium contraction timing ($t_{cra}$) maintained their influence, suggesting their robustness across various scenarios. Such adjustments underscore the importance of developing models that can deliver reliable predictions even when certain data cannot be practically or safely obtained.\\
\\
The exclusions and adjustments, while enhancing clinical safety and applicability, introduce a need for careful validation of the models when applied to individual patient conditions. For Model 1, ensuring the accuracy of non-invasively measured parameters like systemic artery pressure becomes crucial since inaccuracies in these measurements could lead to significant errors in predictions. This careful validation must confirm that the models remain accurate and reliable even when relying solely on non-invasive measurements. For Model 2, the extensive use of non-invasive measures necessitates a comprehensive validation process to ensure that the model can adequately capture the nuanced dynamics of a more complex cardiovascular system. The exclusion of certain parameters from sensitivity analysis highlights the model's adaptability but also points to the necessity of confirming that these adaptations do not compromise the model's predictive accuracy or clinical relevance.

\subsection{Future improvement}

We might consider using wider and more realistic ranges for model distribution instead of 10\% and 1\%, respectively, as the outcomes are predominantly influenced by the first-order effects of the input parameters within this range, indicating the interactions between parameters are relatively low. Future iterations of the model could benefit from incorporating a broader range of parameter variations, possibly capturing more complex interactions that occur under extreme physiological conditions. Extending the parameter range would likely expose higher-order interactions, providing a more accurate depiction of cardiovascular dynamics under stress or disease states. This would also allow us to look into S2 indices, which highlight the interaction between pairs of parameters.\\
\\
We might also consider incorporating more patient-specific data rather than relying solely on general data where each parameter value falls within the normal range. Given the variability in cardiovascular diseases, it's essential to include some extreme values in the sensitivity analysis to better reflect diverse clinical conditions. Additionally, using patient-specific data would allow the model to be more finely adapted to individual needs, thereby enhancing its predictive accuracy and clinical utility. This method is in line with the emerging trends in personalised medicine, which emphasises customising treatments to the unique characteristics of each patient.\\
\\
In this scenario, it would be beneficial to analyse the convergence patterns of each parameter using gradient-based or local sensitivity analysis. This approach would help identify critical points or thresholds around specific personalised values where parameter impacts might shift significantly. By examining how model responses change with small variations in parameters, we can pinpoint precise value ranges that are crucial for model stability and accuracy. Extending this analysis further, we could explore the implications of these critical points for model predictions, particularly in terms of their relevance to different clinical conditions and therapeutic interventions.

\bibliographystyle{IEEEtran}
\bibliography{sample.bib}

\end{document}